\documentclass[final,a4paper,11pt]{article}
\pdfoutput=1
\usepackage[utf8]{inputenc}
\usepackage[T1]{fontenc}

\usepackage{jheppub}
\usepackage{natbib}
\usepackage{subcaption}

\usepackage{amsmath, amsfonts, amssymb}
\usepackage{mathtools}
\usepackage{physics}

\usepackage{graphicx}

\usepackage{shuffle} %% define shuffle symbol notation
\usepackage{slashed} %% Feynman slash notation

\usepackage{cprotect}

\usepackage{xcolor}
\hypersetup{%
    colorlinks=true,%
    allcolors={blue!60!black}%
}

\usepackage[disable]{easy-todo}
\usepackage{cleveref}

\usepackage{booktabs}
\usepackage{array}

\usepackage{listings}
\newcolumntype{L}{>{$}l<{$}}
\newcolumntype{T}{>{\centering\arraybackslash}m{4.5cm}}

\newcommand{\pa}{\partial}
\newcommand{\p}[1]{(\ref{#1})}
\newcommand{\ep}{\epsilon}

\newcommand{\ii}{\,\mathrm{i}\,}

\DeclareMathOperator{\Li}{Li}
\DeclareMathOperator{\sgn}{sgn}

\graphicspath{{figures/}}

\begin{document}

\title{
    Pentagon Functions for Scattering of Five Massless Particles
}

\preprint{MPP-2020-171}

\author[a]{D.~Chicherin}
\emailAdd{chicheri@mpp.mpg.de}
\author[a]{V.~Sotnikov}
\emailAdd{sotnikov@mpp.mpg.de}

\affiliation[a]{Max Planck Institute for Physics (Werner Heisenberg Institute), D--80805 Munich, Germany}

\abstract{
  We complete the analytic calculation of the full set of two-loop Feynman integrals required for computation of massless five-particle scattering amplitudes.
  We employ the method of canonical differential equations to construct a minimal basis set of transcendental functions, \emph{pentagon functions},
  which is sufficient to express all planar and nonplanar massless five-point two-loop Feynman integrals in the whole physical phase space.
  We find analytic expressions for pentagon functions which are manifestly free of unphysical branch cuts.
  We present a public library for numerical evaluation of pentagon functions suitable for immediate phenomenological applications.
}

\maketitle

\listoftodos

\section{Introduction}
Scattering amplitudes are among the central objects of interest in quantum field theories (QFT). 
On the one hand, they are the building blocks for scattering cross sections, which are the crucial theoretical input for phenomenological studies of high-energy particle collisions.
On the other hand, they exhibit intriguing mathematical properties which provide us an opportunity to understand fundamental structure of QFTs.
The coupling constants of the Standard Model are small in the high-energy regime, which
implies that the scattering amplitudes can be consistently approximated by their perturbative expansion.
Beyond the leading order in the expansion, the amplitudes are represented as sums of Feynman integrals with increasing number of loops.
Only one-loop integrals are known for arbitrary scattering processes \cite{vanHameren:2010cp,Denner:2016kdg,Carrazza:2016gav}.
Evaluation of Feynman integrals with two or more loops is an open problem and an active area of studies in theoretical physics and mathematics.
While two-loop integrals for many $2\to 2$ scattering processes have been already obtained (for a recent review see \cite{Amoroso:2020lgh,Heinrich:2020ybq}),
$2\to 3$ processes are on the current frontier of research.
Massless Feynman integrals play a special role in QFT.
The most abundantly produced particles in hadron collisions are the partons of quantum chromodynamics (QCD): gluons and quarks.
Both can be treated as massless at sufficiently high energies.
On a formal side, mathematical structure of QFT is more transparent in the absence of (spontaneously) broken symmetries.

A large number of Feynman integrals contributing to scattering amplitudes can be reduced to a smaller set of \emph{master} integrals
with the help of integration-by-parts identities \cite{Chetyrkin:1981qh}. 
It is a formidable challenge for multi-scale processes, and a number of novel ideas and algorithms has been developed to tackle integral reduction of five-particle processes
\cite{Chawdhry:2018awn,Peraro:2016wsq,Peraro:2019svx,Bendle:2019csk,Wang:2019mnn,Klappert:2019emp,Klappert:2020aqs,Klappert:2020nbg,Boehm:2018fpv,Ita:2015tya,Abreu:2018zmy,Guan:2019bcx}.
Thanks to the advances in integral reduction and functional reconstruction techniques \cite{vonManteuffel:2014ixa,Peraro:2016wsq,Peraro:2019svx},
we have witnessed a tremendous progress in calculation of two-loop five-particle amplitudes.
All planar five-point QCD helicity amplitudes  have been obtained in \cite{Badger:2018gip,Badger:2017jhb,Abreu:2017hqn,Abreu:2018zmy,Badger:2018enw,Abreu:2018jgq,Abreu:2019odu}.
The first results for non-planar five-point amplitudes were obtained in $\mathcal{N}=4$ super-Yang-Mills theory \cite{Chicherin:2018yne,Abreu:2018aqd} and in $\mathcal{N}=8$ supergravity \cite{Chicherin:2019xeg,Abreu:2019rpt},
followed by the full-color five-gluon amplitude with all positive helicities \cite{Badger:2019djh}. Also the full-color six-gluon all-plus helicity amplitude was obtained in \cite{Dalgleish:2020mof}. 
Important progress has been made in evaluation of five-point amplitudes and integrals with one massive leg \cite{Abreu:2020jxa,Hartanto:2019uvl,Papadopoulos:2019iam}.
The first cross section computation of a $2\to 3$ process was carried out in \cite{Chawdhry:2019bji}, where the planar two-loop amplitudes for 
the $q\bar{q}\to \gamma\gamma\gamma$ process were evaluated on a small set of phase space points to construct an interpolating function.

The master integrals for massless five-point scattering processes have been a subject of extensive studies in recent years.
The method of differential equations (DE) \cite{Kotikov:1991pm,Kotikov:1990kg,Remiddi:1997ny,Bern:1993kr,Gehrmann:1999as}
in their canonical form \cite{Henn:2013pwa,Meyer:2017joq,Dlapa:2020cwj,Henn:2020lye,Chen:2020uyk},
and systematic understanding of the transcendental functions appearing in calculations of multi-scale Feynman integrals \cite{Goncharov:2001iea,Goncharov:2010jf,Brown:2011,Chen:1977,Duhr:2011zq}
proved to be indispensable to obtain analytic results for five-point massless master integrals for planar \cite{Gehrmann:2015bfy,Papadopoulos:2015jft,Gehrmann:2018yef}
and non-planar \cite{Chicherin:2018mue,Abreu:2018rcw,Abreu:2018aqd,Chicherin:2018old,Badger:2019djh} topologies.
Differential equations in canonical form provide a natural framework for expressing master integrals
in terms of functions of \emph{uniform transcendental} (UT) weight order by order in the dimensional regulator.
It is advantageous, both for analyzing analytic structure of scattering amplitudes and for their efficient numerical evaluations, 
to have a good grasp on the analytic understanding of the relevant space of transcendental function.
Finding a minimal set of transcendental functions that is sufficient to express all master integrals is essential for deriving compact analytic representations of scattering amplitudes and studying their asymptotic behavior in singular limits (soft, collinear, high-energy, etc.). Successful applications of modern semi-numerical approaches to analytic reconstruction of amplitudes \cite{Peraro:2016wsq,Peraro:2019svx,Abreu:2018zmy,Abreu:2019odu,Badger:2018enw,Badger:2017jhb} rely to a large extent on the knowledge of this set.
At the same time, the representation of amplitudes in terms of a minimal set of transcendental functions achieves the efficiency required in phenomenological applications, where they have to be numerically evaluated on huge samples of phase space points.
In the context of five-particle scattering we refer to this basis set as \emph{pentagon functions}.

For planar massless five-point integrals, a set of pentagon functions was constructed in \cite{Gehrmann:2018yef} and a reference implementation of their numerical evaluation was provided.
For the scattering processes involving solely QCD partons, only planar Feynman integrals contribute in the leading-color approximation.
Obtaining complete NNLO predictions and assessing the accuracy of this approximation requires calculation of amplitudes involving non-planar Feynman integrals.
The scattering processes with photons in the final state also involve non-planar contributions originating from closed fermion loops, which in general cannot be considered subleading.
The knowledge of the full set of pentagon functions is vital for obtaining scattering amplitudes for these classes of processes.
In this work, we construct a basis set of pentagon functions that is sufficient to express all planar and \emph{non-planar} massless five-point two-loop master integrals  in any physical scattering channel.
We consider series expansion of the master integrals in the dimensional regulator up to the order sufficient for calculation of two-loop hard functions and next-to-next-to-leading order (NNLO) cross sections.
The pentagon functions manifestly posses only physical branch cuts.
In particular, they are branch-cut-free in the whole physical phase space and do not require analytic continuation.
We find explicit representations of pentagon functions which admit efficient and stable numerical evaluations, and we implement the latter in a \texttt{C++} library (\cref{sec:cpplib}).
In addition to extending the set of pentagon functions to the non-planar sector, at the same time we reconsider the analysis of the planar sector carried out in \cite{Gehrmann:2018yef}.
The planar subset of pentagon functions presented in the current work is explicitly closed under permutations of external momenta and involves a much smaller set of transcendental constants.
Furthermore, the numerical evaluation of the pentagon functions is significantly improved both in speed and precision.
Thus, for the first time, we provide an implementation that is immediately applicable to computations of NNLO cross sections of any scattering process involving five massless particles.

To find a minimal set of pentagon functions, we follow a constructive approach, which relies almost entirely on the information contained in the canonical differential equations
\cite{Gehrmann:2018yef,Chicherin:2018mue,Abreu:2018aqd,Abreu:2018rcw,Chicherin:2018old,Gehrmann:2000zt,Gehrmann:2001ck}.
We consider the DEs for the planar pentagon-box, hexagon-box, double pentagon, and the one-loop pentagon integral topologies (see \cref{sec:MIbasis}) in all $5!$ permutations of external momenta.
We solve each DE in terms of \emph{iterated integrals} \cite{Chen:1977} with an initial point $X_0$ in the physical scattering region (\cref{sec:DE}).
To completely fix the solutions of DEs one needs to provide the initial values --- values of the master integrals at $X_0$.
Building on the results of \cite{Badger:2019djh,Caron-Huot:2020vlo}, 
we obtain a complete set of the initial values of all DEs at $X_0$ from the requirement of absence of unphysical singularities
and identify a generating set of 19 algebraically-independent transcendental constants.
We then employ the shuffle algebra of iterated integrals (see e.g.\ \cite{Brown:2011}) to find a set of linear-independent irreducible iterated integrals up to transcendental weight four in \cref{sec:classif}.
We evaluate the iterated integrations up to weight two in terms of logarithms and dilogarithms,
and we derive one-fold integral representations \cite{Caron-Huot:2014lda, Gehrmann:2018yef} for the iterated integrals of weight three and weight four.
In this way, we find expressions for all master integrals in any scattering channel sidestepping a difficult problem of analytic continuation.
The obtained analytic expressions allow us to perform a detailed analysis of their behavior in singular limits. As an example,
in \cref{sec:delta0} we investigate the behavior of pentagon functions on boundaries of the physical phase space where all five momenta belong to a three-dimensional subspace,
but none of the external momenta are soft or collinear.
Confirming the observation of \cite{Caron-Huot:2020vlo}, we find that certain weight three and weight four pentagon functions contributing to non-planar master integrals are divergent on these boundaries.

All results of the paper are made available through data files and can be explored with the \texttt{Mathematica} package presented in \cref{sec:numerical}.
We elaborate on the implementation details of pentagon function numerical evaluation by the \texttt{C++} library and demonstrate its performance.
In \cref{sec:validation}, we discuss validation of our results, and we conclude in \cref{sec:conclusions}.

\section{Kinematics}
\label{sec:kinematics}
We study the scattering of five massless particles in four-dimensional Minkowski space-time. 
The particles momenta $p_i$ are subject to momentum conservation $\sum_{i=1}^{5} p_i = 0 $, and on-shell conditions $p_i^2 = 0$.
We parametrize points $X$ of the physical phase space as
\begin{equation} \label{eq:kinematics}
    X =  \left(v_1,\, v_2,\, v_3,\, v_4,\, v_5;\, \ep_5\right) ~\coloneqq~ \left(s_{12},\,s_{23},\,s_{34},\,s_{45},\,s_{15};~4\ii\varepsilon(p_1,p_2,p_3,p_4)\right), \\
\end{equation}
where $s_{ij} := (p_i + p_j)^2$ are the Mandelstam invariants, and $\varepsilon(\boldsymbol\cdot,\boldsymbol\cdot,\boldsymbol\cdot,\boldsymbol\cdot )$ is the fully anti-symmetric Levi-Civita symbol.
The parity-odd invariant $\ep_5$ is related to the determinant of the Gram matrix $G(p_1,p_2,p_3,p_4)$ as 
\begin{equation}
  \Delta  \coloneqq  \qty(\ep_5)^2 =  \det G(p_1,p_2,p_3,p_4) = \det\qty{2~p_i \cdot p_j}, \quad i,j \in \{1,2,3,4\}. \label{delta2ep5}
\end{equation}
In the physical phase space $\Delta<0$ \cite{Byers:1964ryc}, so it is convenient to define 
\begin{equation}
\delta \coloneqq \Im(\sqrt{\Delta}) = \Im(\ep_5),
\end{equation}
such that $\delta \in \mathbb{R}$.
It is worth noting that although $\abs{\delta}$ is not algebraically independent from $v_i$, the \emph{sign} of $\delta$ is necessary
to fully specify a point in the physical phase space.

Depending on the problem at hand, one can choose different parametrizations of the scattering kinematics (see e.g.\ appendix of \cite{Gehrmann:2018yef}).
Our choice of the parametrization in \cref{eq:kinematics} is motivated by the fact that $X$ transforms linearly upon permutations of momenta $p_i$.

\section{Integral Topologies}
\label{sec:MIbasis}
\begin{figure}[th]
  \centering
  \begin{subfigure}[t]{0.4\textwidth}
    \includegraphics[width=\textwidth]{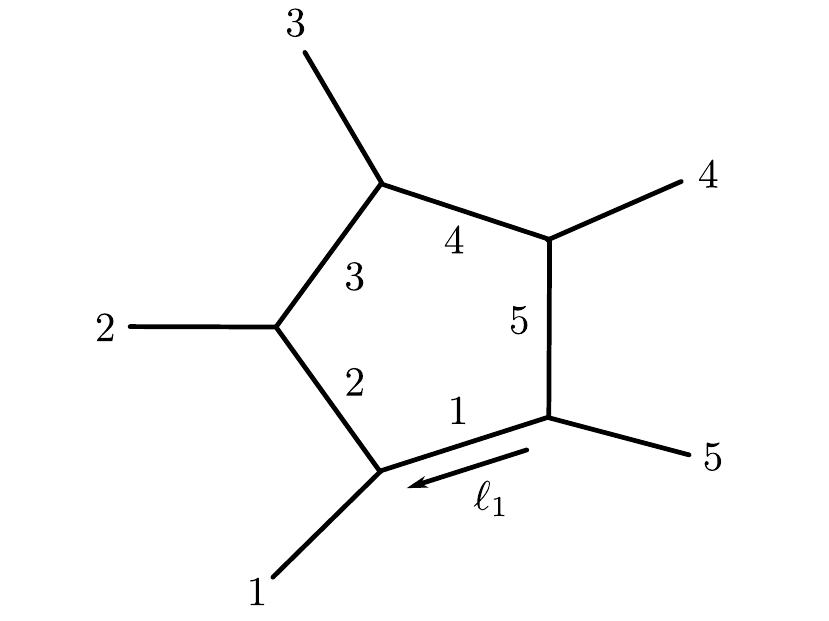}
    \caption{pentagon}
    \label{fig:pentagon}
  \end{subfigure}
  ~
  \begin{subfigure}[t]{0.4\textwidth}
    \includegraphics[width=\textwidth]{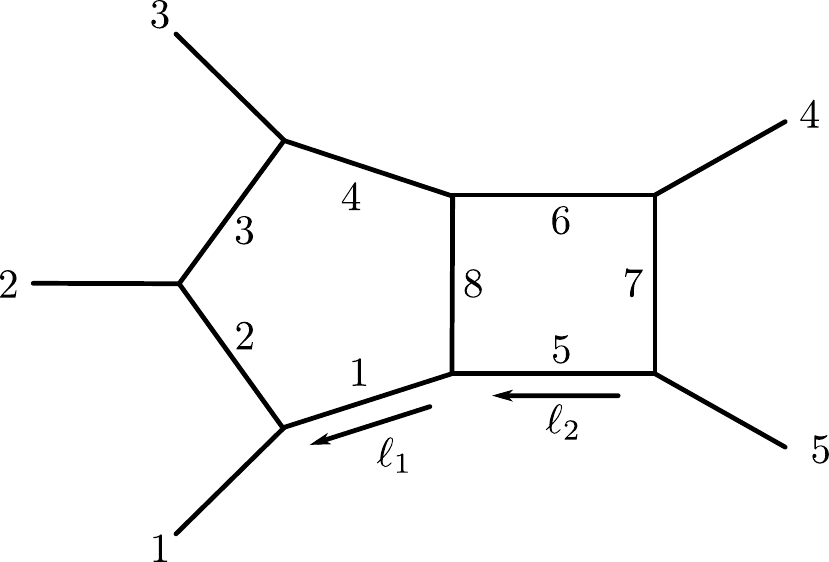}
    \caption{planar pentagon-box}
    \label{fig:pentabox}
  \end{subfigure}

  \begin{subfigure}[t]{0.4\textwidth}
    \includegraphics[width=\textwidth]{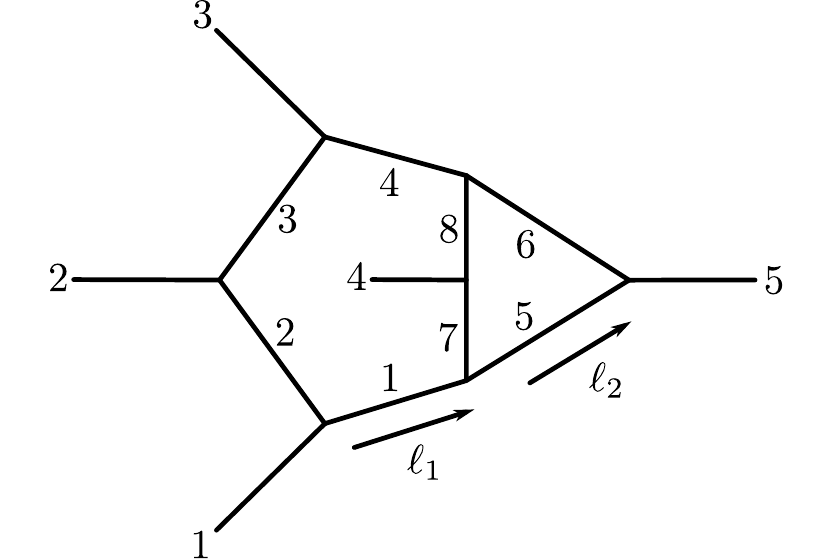}
    \caption{non-planar hexagon-box}
    \label{fig:hexabox}
  \end{subfigure}
  ~
  \begin{subfigure}[t]{0.4\textwidth}
    \includegraphics[width=\textwidth]{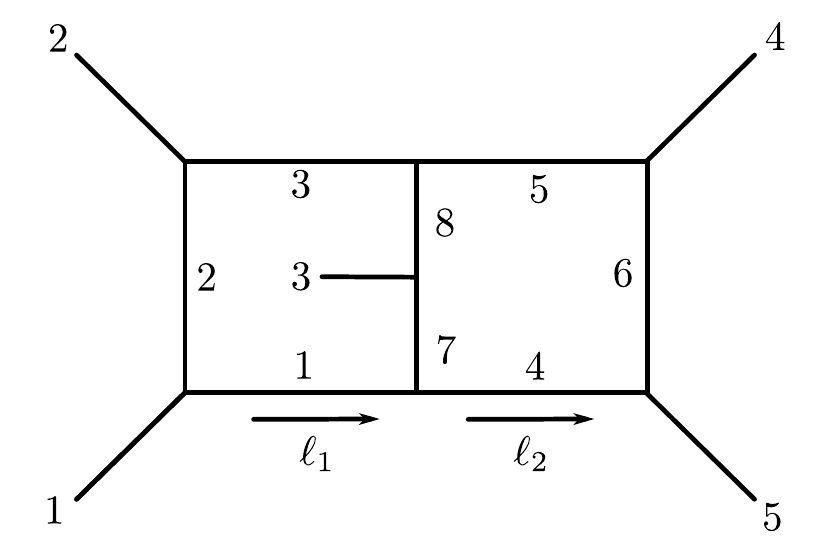}
    \caption{non-planar double pentagon}
    \label{fig:doublepentaong}
  \end{subfigure}
  
  \caption{
    All integral topologies with the maximal number of denominators from each integral family considered in this work.
    Momenta and propagator indices are shown for the standard permutation $\sigma_0$.
    All lines are massless, and all external momenta are incoming.
    The numbering of the denominators corresponds to \cref{eq:denominators}.
  }
  \label{fig:integral_families}
\end{figure}

We consider all Feynman integral topologies required in computation of two-loop scattering amplitudes of five massless particles. There are four topologies, see \cref{fig:integral_families}. Their legs are decorated with particle labels what we call the topology {\it permutation}. To regularize the divergences of loop integrals we employ dimensional regularization and extend the integration measure to $D=4-2 \epsilon$ dimensions. We define the integral families $G_{\tau,\sigma}$ for each topology $\tau$ in permutation $\sigma$ as 
\begin{equation}\label{eq:integrals_def}
  G_{\tau,\sigma}\left[\va*{a}\right] \coloneqq e^{\epsilon L_\tau \gamma_E} \qty(\mu^{2})^{\epsilon L_\tau}
  \int \qty( \prod_{i=1}^{L_\tau} \frac{\dd[D]{\ell_i}}{\ii\pi^{\frac{D}{2}}} )  ~ \frac{1}{\va*{D}_{\tau,\sigma}^{\va*{a}}} ,
   \qquad \va*{D}_{\tau,\sigma}^{\va*{a}} = \prod_i D^{a_i}_{\tau,\sigma\; i}
\end{equation}
where $\gamma_E$ is the Euler-Mascheroni constant%
\footnote{
In this normalization $\gamma_ E$ does not appear in the expressions for integrals.
},
$L_\tau$ is a number of loops in topology $\tau$,
$\va*{D}_{\tau,\sigma}$ is an ordered set of inverse propagators of integral topology $\tau$ in permutation $\sigma$,
the exponents $a_i \in \mathbb{Z}$ for $i \in [1,8]$ and $a_i \in \mathbb{Z}_{\leq 0}$ for $i \in [9,11]$,
and $\mu^2$ is an arbitrary regularization scale, which preserves the integer dimensions of the integrals.
In this paper, we choose the units of energy such that $\mu=1$.
The explicit dependence on the regularization scale can then be restored by the dimensional analysis.

For each of the four integral topologies we choose the standard permutation $\sigma_0 = (1,2,3,4,5)$,
and define the sets $\va*{D}_{\tau,\sigma_0}$ as
\begin{equation} \label{eq:denominators}
  \begin{tabular}{LLLLL}
    \toprule
        & \va*{D}_{\text{a},\sigma_0}  & \va*{D}_{\text{b},\sigma_0}  & \va*{D}_{\text{c},\sigma_0} & \va*{D}_{\text{d},\sigma_0} \\
    \midrule
    1   & \left(\ell_1\right)^2              & \left(\ell_1\right)^2               & \left(\ell_1\right)^2                 & \left(\ell_1\right)^2              \\
    2   & \left(\ell_1 + p_1\right)^2        & \left(\ell_1 + p_1\right)^2         & \left(\ell_1 - p_1\right)^2           & \left(\ell_1 - p_1\right)^2        \\
    3   & \left(\ell_1 + p_1 + p_2 \right)^2 & \left(\ell_1 + p_1 + p_2 \right)^2  & \left(\ell_1 - p_1 - p_2 \right)^2    & \left(\ell_1 - p_1 - p_2 \right)^2 \\
    4   & \left(\ell_1 - p_4 - p_5 \right)^2 & \left(\ell_1 - p_4 - p_5 \right)^2  & \left(\ell_1 + p_4 + p_5 \right)^2    & \left(\ell_2\right)^2 \\
    5   & \left(\ell_1 - p_5 \right)^2       & \left(\ell_2\right)^2               & \left(\ell_2\right)^2                 & \left(\ell_2 + p_4 + p_5\right)^2              \\
    6   &                                    & \left(\ell_2 - p_4 - p_5 \right)^2  & \left(\ell_2 + p_5 \right)^2          & \left(\ell_2 + p_5 \right)^2 \\
    7   &                                    & \left(\ell_2 - p_5 \right)^2        & \left(\ell_1 - \ell_2 \right)^2       & \left(\ell_1 - \ell_2 \right)^2       \\
    8   &                                    & \left(\ell_1 - \ell_2 \right)^2     & \left(\ell_1 - \ell_2 + p_4 \right)^2 & \left(\ell_1 - \ell_2 + p_3 \right)^2    \\
    \midrule
    9   &                                    & \left(\ell_1 - p_5 \right)^2        & \left(\ell_2 - p_1 \right)^2          & \left(\ell_1 + p_5 \right)^2       \\
    10  &                                    & \left(\ell_2 + p_1 \right)^2        & \left(\ell_2 - p_1 - p_2 \right)^2    & \left(\ell_2 - p_1 \right)^2       \\
    11  &                                    & \left(\ell_2 + p_1 + p_2 \right)^2  & \left(\ell_2 + p_4 + p_5 \right)^2    & \left(\ell_2 - p_1 - p_2 \right)^2 \\
    \bottomrule
  \end{tabular}
\end{equation}
This choice is illustrated in \cref{fig:integral_families},
where we show the \emph{top topology} (the topology with maximal number of denominators) for each integral family in the standard permutation $\sigma_0$.
The denominator-variable sets in other permutations $\sigma = (\sigma_1,\ldots{},\sigma_5) \in \mathcal{S}_5$ are generated by the action of the symmetric group $\mathcal{S}_5$
on the set of external momenta $p_i$,
\begin{equation} \label{eq:perm-pi}
  \sigma(p_i) = p_{\sigma(i)} = p_{\sigma_i}.
\end{equation}

The Feynman integrals from each family $G_{\tau,\sigma}$ in \cref{eq:integrals_def} form a linear vector space.
For each $G_{\tau,\sigma}$ (separately) we choose a set of basis elements, which are independent under 
the linear relations generated from integration-by-parts identities \cite{Chetyrkin:1981qh}.
We refer to these sets as \emph{master integrals}.
One can further decrease the number of master integrals in $\cup_{\tau,\sigma}G_{\tau,\sigma}$
by identifying integrals among different topologies and permutations.
However, as we explain in \cref{sec:classif}, we find that it is more convenient to resolve these relations together with the functional relations while constructing a basis of transcendental functions.

A choice of a basis in the vector space $G_{\tau,\sigma}$ is in general arbitrary.
Frequently it is specified by an ordering relation on the set of exponents $a_i$ in \cref{eq:integrals_def} \cite{Laporta:2001dd}  (see also \cite{Smirnov:2020quc,Usovitsch:2020jrk}).
However, it was observed in \cite{Henn:2013pwa} that certain integrals have particularly nice properties.
Following the approach of \cite{Chicherin:2018old, Chicherin:2018mue}, we choose the master integrals with constant leading singularities in $D$ dimensions.
The differential equation for such integrals can be cast into the canonical form \cite{Henn:2013pwa},
and the integrals can be expressed as $\mathbb{Q}$-linear combinations of pure functions with uniform transcendentality (UT).  In the following we call them \emph{UT master integrals}. We discuss the construction and the solution of the differential equations in the next section.

\section{Differential Equations}
\label{sec:DE}
To obtain analytic expressions for all master integrals from integral families defined in \cref{eq:integrals_def},  
we construct the corresponding differential equations \cite{Kotikov:1991pm,Kotikov:1990kg,Remiddi:1997ny,Bern:1993kr,Gehrmann:1999as} in the canonical form \cite{Henn:2013pwa}.
The differential equations (DE) for all integral topologies in \cref{fig:integral_families} have been extensively studied in literature.
The canonical form of differential equations has been obtained in \cite{Gehrmann:2018yef,Chicherin:2018mue,Abreu:2018aqd,Abreu:2018rcw,Chicherin:2018old}. 
The sub-topologies of \cref{fig:integral_families} with less than five external momenta were also studied in \cite{Gehrmann:2000zt,Gehrmann:2001ck}.

For the double-pentagon topology, we directly use the canonical DE of \cite{Chicherin:2018old}.
For hexagon-box, planar pentagon-box, and one-loop pentagon topologies we repeat the analysis of \cite{Chicherin:2018old}
to find master integrals with unit leading singularities in $D$-dimensions. Their four-dimensional integrands have $\dd{\log}$ form. 

In this section we provide details of the construction and integration of DEs, which
are necessary for the construction of a basis of transcendental functions in \cref{secStrategy}.

\subsection{Construction of canonical differential equations}
\label{sec:constructionDE}

We would like to find the analytic expressions for all integral topologies in \cref{fig:integral_families} in all $5!$ permutations.
One can think of two different approaches to constructing DE solutions for all permutations.
In the first approach, one would consider a single permutation
of each topology, e.g.\ the one depicted in \cref{fig:integral_families}, solve it analytically by
the method of differential equations, and then obtain analytic expressions for
all other permutations of the topology by means of  analytic continuation.
The latter is a highly nontrivial task for integrals depending on many scales.
In particular, some of the non-planar master integrals develop discontinuities and even divergences 
inside the physical region on subvarities of $\Delta = 0$ without collinear or soft momenta \cite{Caron-Huot:2020vlo};
we discuss this in \cref{sec:delta0}.
In this paper we follow an alternative approach which was advocated in \cite{Gehrmann:2018yef}.
We work simultaneously with all $5!$ permutations of each topology in \cref{fig:integral_families} and consider canonical differential equations (DE) for each of them 
\begin{subequations}\label{difeq}
  \begin{align}
    \dd \vec f_{\tau,\sigma} & = \epsilon \, \dd \tilde{A}_{\tau,\sigma} \vec f_{\tau,\sigma}, \\
    \dd \tilde A_{\tau,\sigma} & = \sum_{i = 1}^{31} a^{(i)}_{\tau,\sigma} \, \dd\log W_i \,
  \end{align}
\end{subequations}
where $\vec f_{\tau,\sigma}$ is a vector of UT master integrals of topology $\tau$ taken in permutation $\sigma$. Entries of the matrices $a^{(i)}_{\tau,\sigma}$ are rational constants, and $\{ W_i\}_{i=1}^{31}$ are letters of the pentagon alphabet \cite{Chicherin:2017dob}.
The letters are algebraic functions of the Mandelstam variables.
We review the pentagon alphabet in \cref{App_alph}.
The matrices of the DE in permutation $\sigma$ are related to the DE in the standard permutation $\sigma_0$ as follows,    
\begin{align}
\dd \tilde A_{\tau,\sigma} = \sigma(\dd \tilde{A}_{\tau, \sigma_0}) = \sum_{i=1}^{31} a^{(i)}_{\tau,\sigma_0} \dd\log \sigma(W_i) \,,
\end{align}
where $\sigma$ permutes the external momenta according to \p{eq:perm-pi}.
The pentagon alphabet is closed under ${\cal S}_5$ permutations of the external momenta,
and the letters of the alphabet have simple transformation properties.
In particular, the set of $\dd\log W_i$ integration kernels forms a linear representation of ${\cal S}_5$.
We refer to \cref{App_alph} for details.
Thus, we find canonical DEs for all permutations starting with the canonical DE for a single permutation.

\begin{table}[t]
\begin{center}
  \begin{tabular}{Tcccc}
\toprule
Topology $\tau$ & pentagon & pentagon-box & hexagon-box & double pentagon \\
\midrule
\# master integrals & 10|1 & 53|8 & 62|11 & 88|20 \\
\# master integrals on top topology & 0|1 & 1|2 & 1|2 & 3|6 \\
\bottomrule
\end{tabular}
\caption{Number of parity-even|parity-odd master integrals (a single permutation) in the four integral topologies defined in \cref{fig:integral_families}.}
\label{tab_count_mast}
\end{center}
\end{table}

The master integrals $\vec f_{\tau,\sigma}$ are Lorentz-invariant functions of the five momenta, and by the definition of the topology permutation we have 
\begin{align}
  \vec f_{\tau,\sigma} (X) =  \vec f_{\tau,\sigma_0}(\sigma X) \,, \label{fperm}
\end{align}
where the action of $\sigma \in \mathcal{S}_5$ on the kinematic point $X$ (see \cref{eq:kinematics})
is induced from the action of $\sigma$ on momenta $p_i$ by \cref{eq:perm-pi}.
In addition, for arbitrary $\eta\in \mathcal{S}_5$, the following relation holds,
\begin{equation} \label{eq:perm-composition}
  \vec f_{\tau,\sigma} ( \eta X) =  \vec f_{\tau,\, \eta\sigma}(X).
\end{equation}

The UT master integrals in the standard permutation are related to the integrals from \p{eq:integrals_def} by linear transformations
\begin{align}
  \vec f_{\tau}  \coloneqq \vec f_{\tau,\sigma_0} = \sum_{\va*{a}} \vec T_{\tau} [\va*{a}] \, G_\tau[\va*{a}]  \,.
\label{changebasis}
\end{align}
The UT master integrals $\vec f_\tau$ of our basis are split in the parity-even $\vec{f}^{(+)}_{\tau}$ and the parity-odd $\vec{f}^{(-)}_{\tau}$ ones. 

For the parity-even master integrals the transformation coefficients $\vec T^{(+)}_{\tau}$ are rational functions of the Mandelstam invariants,
and for the parity-odd integrals, the coefficients $\vec T^{(-)}_{\tau}$ are in addition proportional to the parity-odd invariant $\ep_5$ (see \cref{eq:kinematics}), 
\begin{subequations}\label{Tdef}
  \begin{align}
    T^{(+)}_{\tau} & \in \mathbb{Q}(v_{1},\ldots,v_{5}) , \\
    T^{(-)}_{\tau} & \in \ep_5 \mathbb{Q}(v_{1},\ldots,v_{5})\,.    
  \end{align}
\end{subequations}
The number of parity-even and parity-odd master integrals (i.e.\ dimensions of $\vec f_{\tau}^{(\pm)}$) are given in \cref{tab_count_mast}.
We provide explicit transformations \eqref{changebasis} in ancillary files (see \cref{sec:math_package}).
 
In the next section, we solve simultaneously all $4 \times 5!$ differential equations \p{difeq}.

\subsection{Integrating DE and iterated integrals}

To integrate DEs \p{difeq} order-by-order in $\epsilon$, we define $\epsilon$-expansions of the UT master integrals as
\begin{subequations}
  \begin{align}
  & \vec f_{\tau,\sigma} = \sum_{w \geq 0} \frac{1}{\epsilon^{4-w}}  \, \vec f^{(w)}_{\tau,\sigma} \;,
      \qquad \tau = \text{penta-box}\, (b), \; \text{hexa-box}\, (c), \; \text{double pentagon}\, (d) \label{UTint_eps_exp} \\
  & \vec f_{\tau,\sigma} = \sum_{w \geq 0} \frac{1}{\epsilon^{2-w}}  \, \vec f^{(w)}_{\tau,\sigma} \;,\qquad \tau = \text{pentagon}\, (a)
  \label{UTint1loop_eps_exp}
  \end{align}
\end{subequations}
where $\vec f^{(w)}_{\tau,\sigma}$ are of uniform transcendental weight $w$. 
The $\epsilon$-expansion of the two-loop master integrals starts with $\frac{1}{\epsilon^4}$ pole, and with $\frac{1}{\epsilon^2}$ pole for one-loop pentagons, i.e.\ 
the soft-collinear pole $\frac{1}{\epsilon^2}$ per loop order.
We omit in the following the topology and permutation labels $\tau,\sigma$ to avoid bulky notations. 

Let us denote a point of the kinematic space by $X$ \p{eq:kinematics}. 
We specify it by the set of five adjacent Mandelstam invariants and the sign of $\epsilon_5$. 
We choose an {\it initial point} $X_0$ and integrate the DE along a path $\gamma$  connecting $X_0$ and $X$. 
Thus we express weight-$w$ solutions at an arbitrary kinematic point $X$ as iterated integrals of the {\it initial values} $\vec f^{(w')}(X_0)$, i.e.\ 
solutions with $w' \leq w$ evaluated at $X = X_0$,
\begin{align}
\vec f^{(w)}(X) = \sum_{w' = 0}^{w} \; \underbrace{\int_{\gamma} \dd\tilde{A} \ldots \int_{\gamma} \dd\tilde{A}}_{w' \; \text{integrations}} \, \vec f^{(w-w')}(X_0) \,.  \label{solDE}
\end{align}
At weight 0 the previous equation simplifies to $\vec f^{(0)}(X) =  \vec f^{(0)}(X_0)$, i.e.\ it is a constant vector of rational numbers. 
Moreover, the vector is the same for all permutations of the given topology,
\begin{align}
\vec f^{(0)}_{\tau,\sigma}(X) = \vec f^{(0)}_{\tau,\sigma_0}(X_0) \,. \label{f0sol}
\end{align}

\Cref{solDE} can be rewritten explicitly as a linear combination of iterated integrals built upon the pentagon alphabet 
\begin{align}
\vec f^{(w)}(X) = \sum_{w' = 0}^{w} \; \sum_{i_1,\ldots,i_{w'} = 1}^{31} \vec \kappa^{(w-w')}_{i_1,\ldots,i_{w'}} \left.[W_{i_1},\ldots,W_{i_{w'}}]\right._{X_0}\!\!(X) \,. \label{f_it_int_gen}
\end{align}
The coefficients $\kappa^{(w-w')}$ of the linear combination are transcendental constants of weight $w-w'$,
\begin{align}
\kappa^{(w-w')}_{i_1,\ldots,i_{w'}} = a^{(i_1)}a^{(i_2)} \ldots a^{(i_{w'})} \vec{f}^{(w-w')}(X_0) \,. \label{kappa}
\end{align}
Chen iterated integrals \cite{Chen:1977} of weight $w$ along the path $\gamma$ are defined recursively as
\begin{align}
\left.[W_{i_1},\ldots,W_{i_{w}}]\right._{X_0}\!\!(X) = \int_\gamma \dd\log W_{i_w}(X') \, \left.[W_{i_1},\ldots,W_{i_{w-1}}]\right._{X_0}\!\!(X') \label{it_int_def}
\end{align}
with $[]_{X_0} = 1$.
The iterated integrals vanish at the initial point $X = X_0$ by construction, 
\begin{align}
\left.[W_{i_1},\ldots,W_{i_{w}}]\right._{X_0}\!\!(X_0) = 0\,, \quad w >0 \, . 
\end{align}

The DE guaranties that only homotopy invariant, i.e.\ invariant under small deformations of the integration contour $\gamma$,
linear combinations of the iterated integrals are present in the solution \p{f_it_int_gen}.
The iterated integrals are in general multi-valued functions since they pick up a nontrivial monodromy upon integrating around a pole,
i.e.\ zero locus of an alphabet letter.
Thus we have to specify an \emph{analyticity region} $P_0$ within which the iterated integrals are single-valued and real-analytic functions. Then the result of the iterated integration depends only on the end points $X$ and $X_0$ of the integration path $\gamma \subset P_0$.

The iterated integral representation is a powerful tool which enables us to classify the solutions of the DE within the analyticity domain by doing simple algebraic calculations. 
In particular, the iterated integrals satisfy the shuffle algebra relations (see e.g.\ \cite{Brown:2011}),
which specify how to rewrite a product of several iterated integrals as a sum of iterated integrals,
\begin{align}
\left.[W_{i_1},\ldots,W_{i_{w_1}}]\right._{X_0} \!\!(X)\,
\left.[W_{j_1},\ldots,W_{j_{w_2}}]\right._{X_0} \!\!(X) = \sum \left.[W_{k_1},\ldots,W_{k_{w_1+w_2}}]\right._{X_0}\!\!(X) \, \label{shuf}
\end{align}
where we sum over all $\{k_1,\ldots, k_{w_1+w_2}\}$ in the shuffle product $\{i_1,\ldots, i_{w_1}\} \shuffle \{j_1,\ldots, j_{w_2}\}$.
After applying the shuffle algebra relations, all polynomial identities among the functions represented by iterated integrals become linear, and only the trivial combination of the iterated integrals vanishes. 
In this way we take into account the functional relations among the DE solutions.

\subsection{Physical region}
\label{sect_phys_reg}

As we have already mentioned, the iterated integrals \p{f_it_int_gen} are not single-valued and real-analytic at any kinematic point $X$.
We choose the following analyticity domain   
\begin{align}
  P_0: \quad s_{12},s_{34},s_{35},s_{45} > 0 
  ,\; s_{13},s_{14},s_{15},s_{23},s_{24},s_{25} < 0, \;
  \Delta < 0, \; \delta > 0\,.  \label{s12channel} 
\end{align}
It is a half of the physical $s_{12}$-channel scattering region, i.e.\ $12 \to 345$ scattering process. 
Fixing signs of the Mandelstam invariants implies that the particle energies are positive and scattering angles are real.
$\Delta <0$ implies the reality of momenta (see also \cref{eq:kinematics,delta2ep5}). 
In addition, we also fix the branch of the square root $\sqrt{\Delta}$ by the condition $\delta>0$.
The boundaries of $P_0$ corresponding to vanishing of one or several $s_{ij}$ describe the soft/collinear limits. 
The boundary $\Delta = 0$ of $P_0$ lies inside the physical $s_{12}$-channel and splits it into two halves. 
It corresponds to the kinematics with all five momenta lying in a three-dimensional hyperplane. 
Crossing of the $\Delta = 0$ variety separating $\Im(\ep_5) > 0$ and $\Im(\ep_5) < 0$ regions is not innocuous since the master integrals could diverge there 
\cite{Caron-Huot:2020vlo}.

In the following we work strictly inside $P_0$ \p{s12channel} and classify the solutions of the DE \p{f_it_int_gen} only for $X \in P_0$. 
Since we consider all $5!$ permutations of each topology, we can immediately translate our results to any physical scattering region.
This, we provide analytic expressions for the master integrals taken in arbitrary permutation \p{eq:integrals_def} through the whole phase space, see \cref{secAnyChannel}.

To completely define the iterated integrals \p{it_int_def}, we also need to specify an initial point $X_0$ and an integration path $\gamma$.
We choose $X_0$ inside $P_0$ \p{s12channel} as follows,
\begin{align} \label{B}
  X_0 ~=~ (v_1^{0},v_2^{0},v_3^{0}, v_4^{0}, v_5^{0};\,\ep_5^{0})~=~(3,-1,1,1,-1;\,\ii \sqrt{3})\,.
\end{align}
This point is invariant under permutations $\left({\cal S}_2 \times {\cal S}_3\right) / \mathbb{Z}_2$ of 
``incoming'' $\{1,2\}$ and ``outgoing'' $\{3,4,5\}$ particles of the $s_{12}$-scattering channel that preserve $\sgn(\delta)$.

Given an arbitrary point $X \in P_0$, we evaluate the iterated integrals by choosing the path $\gamma$ to be a line segment connecting
$X_0$ and $X$, \cref{eq:kinematics}. We parametrize the segment $\gamma : t \mapsto X(t)$ as
\begin{align}\label{Xt}
  v_i(t) = (1-t) \, v_i^{0} ~+~ t\,v_i, \qquad \ep_5(t) = \sqrt{\Delta(v_i(t))}, \qquad t \in [0,1]\,.
\end{align}
If we are to avoid the problem of analytic continuation,
the integration path $\gamma$ must never leave the analyticity domain $P_0$ \cref{s12channel}, i.e.\ for any $X \in P_0$, $\{ X(t)\}_{0\leq t\leq 1} \subset P_0$ must be satisfied.
To this end, we note that $P_0$ is not a convex\footnote{For example, 
let us consider the following line segment $\gamma$ parametrized by $0 \leq t \leq 1$,
\begin{align}
X(t): \qquad
(v_1,v_2,v_3,v_4,v_5) = \left( 7,-\frac{1}{2},\frac{11}{2},1,-1 \right) + t \left( 0,0,- \frac{47}{10},5,0 \right)
\end{align}
We find $\Delta(X(t = 0)) = -\frac{87}{16}<0$ and $\Delta(X(t = 1)) =
-\frac{231}{100}<0$ so the end points of $\gamma$ belong to
$P_0$ \p{s12channel}. However, at the intermediate point $\Delta(X(t = \frac{1}{2})) =
\frac{1323}{50}> 0$, and the segment does not lie inside the physical $s_{12}$-channel.}.  
Nevertheless, a weaker statement holds:
a line segment connecting $X_0$ \p{B} with an arbitrary point $X\in P_0$ lies entirely inside $P_0$.
We outline the proof in \cref{App_conv}.
Thus, integrating in \p{it_int_def} along the straight lines connecting $X_0$ with any $X \in P_0$ we obtain real-analytic single-valued solutions \p{f_it_int_gen} throughout $P_0$.

\subsection{Initial values}
\label{bc}

In order to be able to integrate DEs \p{difeq}, we need to know $\ep$-expansion of all UT master integrals at the reference point $X_0$ \eqref{B} --- the \emph{initial values} of the DEs.
As we pointed out in \cref{sec:constructionDE},
we would like to trade the problem of analytic continuation of the UT master integrals in the standard permutation $\sigma_0$ to all possible permutations,
for solving the DEs with initial values at $X_0$ in all $5!$ permutations:
\begin{gather*}
\{ \vec f_{\tau,\sigma_0}(X_0)\} \; +  \; \text{analytic continuation of } \vec f_{\tau,\sigma_0}(X) \text{ to any physical $s_{ij}$-channel}  \\
  \Big\Updownarrow\\
   \{ \vec f_{\tau,\sigma}(X_0) \}_{\sigma \in {\cal S}_5}
\end{gather*}

We restrict our consideration to weight $w \leq 4$ initial values, i.e.\
we truncate the $\ep$-expansion \p{UTint_eps_exp} of the two-loop topologies at the finite part,
and at ${\cal O}(\ep^2)$ for the one-loop pentagon \p{UTint1loop_eps_exp}. 

Initial values of the DE for two-loop five-point topologies have been extensively studied previously. 
Weight-0 initial values $\vec f^{(0)}$ are rational numbers. 
They are enough to construct symbols \cite{Goncharov:2001iea,Goncharov:2010jf,Brown:2011} of the UT master integrals, and are relatively easy to obtain. 
Calculation of higher weight initial values is much more tedious. 
The planar pentagon-box topology (\cref{fig:pentabox}) has been solved in any physical region in \cite{Gehrmann:2018yef}. 
In \cite{Chicherin:2018mue} the initial values for one permutation of the hexagon-box topology (\cref{fig:hexabox}) were evaluated in the Euclidean region. In \cite{Chicherin:2018old} the initial values in a physical region for one permutation of the double pentagon topology (\cref{fig:doublepentaong}) were presented with 50 digits precision. In calculation of a five-point nonplanar amplitude in \cite{Badger:2019djh}, the initial values at $X_0$ were computed for all permutations of all four topologies in \cref{fig:integral_families} with 200 digit precision, but they were not explicitly reported. The initial values have been fixed by requiring the absence of the unphysical singularities in the DE solutions \p{solDE}, see \cite{Gehrmann:2018yef,Chicherin:2018mue,Henn:2013nsa,Henn:2019rgj}, and special care have been taken owing to the singular behaviour of the nonplanar Feynman integrals at $\Delta = 0$. In the present paper, we publish for the first time the complete set of initial values at $X_0$ (all permutations $\sigma$ of all four topologies $\tau$). 

In \cite{Caron-Huot:2020vlo}, by integrating the DEs, the initial values at $X_0$ were transported to a point $X_{\rm Regge} \in P_0$ in the Regge asymptotic regime. 
In this regime the pentagon alphabet enormously simplifies, which leads to a more simple form of the initial values at $X_{\rm Regge}$ as compared to $X_0$. 
The available numerical precision is enough for fitting to a basis of transcendental constants. 
A small generating set ${\cal S}_{\rm Regge}$ (see table~2 in \cite{Caron-Huot:2020vlo}) of algebraically independent over $\mathbb{Q}$ transcendental constants was identified,
and all initial values at $X_{\rm Regge}$ were written as polynomials $\mathbb{Q}[{\cal S}_{\rm Regge}]$ graded by the transcendentality degree. 
These analytic expressions for the initial values at $X_{\rm Regge}$ were transported back to $X_0$ \p{B} by integrating the DEs \p{difeq}
in terms of \emph{multiple polylogarithms} (MPLs) \cite{Goncharov:2001iea,Goncharov:1998kja} and evaluating the latter with \texttt{GiNaC} \cite{Bauer:2000cp} with $10^4$ digit precision. 
In this way the initial values at $X_0$ have been found with at least $9\cdot 10^3$ digit precision. 
This precision is enough to identify the generating set ${\cal S}_0$ of algebraically independent over $\mathbb{Q}$ transcendental constants,
see \cref{tab_trans_const}, and to fit the initial values at $X_0$ to graded polynomials $\mathbb{Q}[{\cal S}_{0}]$.
The analytic form of the initial values can be found in the data files supplied with the \texttt{Mathematica} package (see \cref{sec:math_package}). 

The set ${\cal S}_{0}$ consists of only 19 transcendental constants, which we classify as follows. We assign  the $\mathbb{Z}_{\geq 0} \times \mathbb{Z}_2$-charge (or grading) to the constants where the first factor refers to the transcendentality degree while the second factor $\mathbb{Z}_2 = \{ +, - \} = \{ \text{\it even}, \text{\it odd} \} $ counts parity. 
Then the weight-$w$ initial values $ f^{(w,+)}_{\tau,\sigma}(X_0)$ of the  parity-even master integrals
are homogeneous polynomials $\mathbb{Q}[{\cal S}_0]_{(w,+)} $, while the initial values $ f^{(w,-)}_{\tau,\sigma}(X_0)$ 
of the  parity-odd master integrals are homogeneous polynomials $\mathbb{Q}[{\cal S}_0]_{(w,-)} $. 
In order to be able to consistently assign the parity to the initial values we have to introduce two copies of $\ii\pi$ of opposite parity, i.e.\
parity-odd $\ii\pi$ and parity-even $\ii\overset{\circ}{\pi}$. 
For example, $\log(3) + \ii\overset{\circ}{\pi}$ carries charge $(1,+)$
and it is an admissible weight-one initial value of a parity-even UT master integral; 
$\pi \overset{\circ}{\pi} + \ii \Im\Li_2\left( e^{\frac{\ii\pi}{3}}\right)$ carries charge $(2,-)$ and it is an admissible weight-two initial value of a parity-odd UT master integral. 
Obviously, $\ii\pi$ and $\ii\overset{\circ}{\pi}$ are numerically identical, and we are allowed to identify $\pi^2 = (\overset{\circ}{\pi})^2$. 
We notice that all parity-odd constants are pure imaginary, and all parity-even constants are real (except for $\ii\overset{\circ}{\pi} $). 
The reality properties of the transcendental constants imply that the initial-values of the parity-even master integrals are real
and the initial-values of the parity-odd UT master integrals are pure imaginary modulo $\ii\overset{\circ}{\pi}$,  
\begin{subequations}\label{ImReInitVals}
  \begin{align}
  & \Im \vec{f}^{(w,+)}_{\tau,\sigma}(X_0) \, \in \, \overset{\circ}{\pi} \,\mathbb{Q}[{\cal S}_0]_{(w-1,+)} \, ,\\
  & \Re \vec{f}^{(w,-)}_{\tau,\sigma}(X_0) \, \in \, \ii\overset{\circ}{\pi} \,\mathbb{Q}[{\cal S}_0]_{(w-1,-)} \,. 
  \end{align}
\end{subequations}
If there existed an Euclidean region for all master integrals $G_\tau[\vec{\mathbf{a}}]$ \p{eq:integrals_def} where they take real values, 
then both parity-even and parity-odd UT master integrals would also be real in that region.
Then, by analytic continuation from the Euclidean region to a physical scattering region the parity-odd integrals $\vec{f}^{(-)}_{\tau}$
would become imaginary up to the contributions from discontinuities, which are proportional to $\ii\pi$.
This would be in agreement with \cref{ImReInitVals}. 
However, non-planar integrals from the double-pentagon topology (\cref{fig:doublepentaong}) do not have an Euclidean region and they are complex-valued everywhere. 
Nevertheless, we find the observed correspondence between the reality properties of the initial values and the parity of the UT master integrals very intriguing.

\begin{table}[t]
  \centering
  \small
  \begin{tabular}{>{$}c<{$}LL}
    \toprule
    \text{Weight} & \text{even $(+)$} & \text{odd $(-)$} \\
    \midrule
    1 & \log(2), \log(3), \ii \overset{\circ}{\pi} & \ii\pi \\ 
    2 & \Li_2\left(\frac{2}{3}\right) &  \ii \Im\Li_2\left(\frac{1}{2}+\frac{\ii\sqrt{3}}{2}\right) \\ 
    3 & \Li_3\left(\frac{2}{3}\right), \Li_3\left(\frac{1}{4}\right), \zeta_3 & \ii \Im\Li_3\left(\frac{\ii}{\sqrt{3}}\right), \ii \Im\Li_3\left(1+ \ii \sqrt{3} \right) \\ 
    4 &  \Li_4\left(\frac{1}{4}\right), \Li_4\left(\frac{1}{3}\right), \Li_4\left(\frac{1}{2}\right), &   \ii \Im\Li_4\left(\frac{1}{2}+\frac{\ii\sqrt{3}}{2}\right), \ii \Im\Li_4\left(\frac{\ii}{\sqrt{3}}\right), \\ 
      &   \Li_4\left(\frac{2}{3}\right), \Li_4\left(\frac{3}{4}\right) & \ii \Im \qty[6 \Li_4 \left( 1- \frac{\ii}{\sqrt{3}}\right) + 6 \Li_4 \left( 1+ \ii \sqrt{3} \right) + 5 \Li_{2,2} \left( \frac{1}{2} + \frac{\ii\sqrt{3}}{2} \right)] \\ 
      \bottomrule
  \end{tabular}
  \caption{The basis ${\cal S}_0$ of algebraically independent over $\mathbb{Q}$ transcendental constants specifiying the initial values $\{ \vec f^{(w)}_{\tau,\sigma}(X_0) \}$ at weights $ w=0,1,\ldots,4$. The elements of ${\cal S}_0$ are charged by $\mathbb{Z}_{\geq 0}$ representing their transcendental weight, and by $\mathbb{Z}_2 = \{ + , - \} = \{ \text{\it even}, \text{\it odd} \}$ representing their parity. We introduced $\ii \overset{\circ}{\pi}$ which equals numerically to $\ii \pi$ but carries even parity. ${\cal S}^{(w,\pm)}_0$ denotes elements of ${\cal S}_0$ with $(w,\pm)$ charge.}\label{tab_trans_const}
\end{table}

\subsection{Parity of the UT master integrals}
\label{sec_parity}

As we discussed in the previous section, the initial values at $X_0$, or equivalently
the transcendental numbers $\kappa$ in \cref{kappa}, obey $\mathbb{Z}_{\geq 0} \times \mathbb{Z}_2$-grading.
The same is true for the iterated integrals \p{it_int_def}.
Indeed, counting of the parity-odd letters (see \cref{App_alph}) in the iterated integral $[W_{i_1},\ldots,W_{i_n}]$ corresponds to the $\mathbb{Z}_2$-grading,
and the number of iterated integrations (weight) corresponds to $\mathbb{Z}_{\geq 0}$-grading.
Both gradings are compatible with the shuffle algebra \p{shuf}.
It then follows that the UT master integrals $\vec f^{(w)}_{\tau,\sigma}(X)$ inherit the $\mathbb{Z}_2$-grading of the initial values and iterated integrals.
This fact allows us to establish an equivalence between the $\mathbb{Z}_2$-grading and the parity of the UT master integrals $\vec{f}^{(\pm)}_{\tau,\sigma}$ implied by the parity-conjugation properties of the coefficients $\vec T_{\tau}$ in definition \p{changebasis}.

We would like to emphasize that this compatibility is not trivial. Indeed, had we not introduced two copies of $\ii\pi$ to represent the initial values,
the equivalence would not hold. Moreover, if taken literally, the parity conjugation maps the initial point $X_0$ \p{B} and the path $\gamma$ of an iterated integral \eqref{it_int_def}
out of the chosen analyticity region $P_0$ into a region with $\delta<0$
in addition to parity conjugation of the $\dd{\log}$-kernels (see \cref{App_alph}).
The relations between parity-conjugated iterated integrals could then be established through analytic continuation.
As discussed in the previous section, our strategy is to avoid analytic continuation in favor of considering momenta permutations of the master integrals, so we do not pursue this approach in what follows.

\section{Classification of Functions}
\label{sec:classif}
Upon integration of DEs \p{difeq} for all four topologies $\tau$, each in $5!$ permutations, we obtain a number of iterated integrals \p{f_it_int_gen} which are not independent. 
We would like to reduce them to a minimal set of functions which are sufficient to express all solutions $\vec f^{(w)}_{\tau,\sigma}$ up to weight $w \leq 4$. 
Before we delve into the classification procedure, it is worth noting that from $(11 + 61 + 73 + 108)\times 5!$ UT master integrals involved in DEs \eqref{difeq}
only $1865$ two-loop and $52$ one-loop UT master integrals are linear-independent under the topological identifications 
among different topologies and permutations of integrals $G_{\tau,\sigma}[\vec{\mathbf{a}}]$ (see \cref{eq:integrals_def}) \cite{Badger:2019djh}.
These relations in the set of UT master integrals are trivialized by solving their DEs in terms of iterated integrals.
For this reason, we do not explicitly implement them in our classification.
In this section, we find the linear-independent solutions at weights $w\leq 4$ and show that their number is smaller than the one obtained from the topological analysis.\footnote{%
It is expected that this redundancy should be lifted by considering higher orders in $\epsilon$-expansion \p{UTint_eps_exp}.
} We then further reduce the set of linear-independent iterated integrals to a smaller set of \emph{irreducible iterated integrals}, i.e.\ 
the ones which cannot be represented as products of lower-weight iterated integrals.
We claim that the latter set is minimal, and we denote it as \emph{pentagon functions}.

\subsection{Classification strategy}

\label{secStrategy}

Let us briefly outline our classification strategy. 
We will assume that the solutions vanish at $X=X_0$, or in other words we consider $\vec f^{(w)}_{\tau,\sigma}(X)- \vec f^{(w)}_{\tau,\sigma}(X_0)$.
We proceed recursively in weight. 
At weights $1 \leq w_1<w$ we have already identified $\{ {\cal I}^{(w_1)}_i(X) \}_{i = 1}^{L_{w_1}}$ the minimal irreducible sets of iterated integrals. 
Let $\{ {\cal I}^{(w)}_1,\ldots,{\cal I}^{(w)}_{N_w} \}$ be the set of all iterated integrals \p{f_it_int_gen} --- $\{ \vec f^{(w)}_{\tau,\sigma}(X) \}^{\tau = (a),(b),(c),(d)}_{\sigma \in {\cal S}_5}$. 
We rewrite them schematically in the following form splitting out the term $I^{(w)}$ with the maximal number of iterated integrations,
\begin{align}
{\cal I}^{(w)} = I^{(w)} + \sum_{w'=1}^{w-1} \sum_a \kappa^{(w-w')}_a \, R^{(w')}_a \,. \label{solsplit}
\end{align}
In other words, $I^{(w)}$ is the symbol of ${\cal I}^{(w)}$; $\kappa^{(w-w')}_a$ are transcendental constants of weight $w-w'$ and $R^{(w')}$ represents $\mathbb{Q}$-linear combination of weight-$w'$ iterated integrals. 
We mod out lower-weight iterated integrals, i.e.\ we apply the symbol map \cite{Goncharov:2001iea,Goncharov:2010jf,Brown:2011} and choose the subset of $M_w \leq N_w$ linear independent 
\begin{align}
I^{(w)}_{i_1},\ldots,I^{(w)}_{i_{M_{w}}} \label{Iset}
\end{align}
in the set $\{ I^{(w)}_{1},\ldots,I^{(w)}_{N_{w}} \}$. 
Then we need to eliminate {\it reducible} iterated integrals from \p{Iset}, i.e.\
iterated integrals which are products of lower weight iterated integrals. 
In order to achieve it, we consider symbols of weight-$w$ products of lower weight iterated integrals,
$\{ {\cal I}^{(w_1)}_i \}_{i = 1}^{L_{w_1}}$ with $w_1<w$, already classified at the previous steps of our procedure, e.g.\
\begin{align}
I^{(w-1)} \times I^{(1)} \;,\; I^{(w-2)} \times I^{(2)} \;,\; 
I^{(w-2)} \times I^{(1)} \times I^{(1)} \;, \ldots \label{prodI}
\end{align}
which are linear independent by induction. 
Using the shuffle algebra \p{shuf} we rewrite them as sums and complement them with the symbols \p{Iset}. 
The resulting set of symbols is overcomplete. Then we choose a basis in their $\mathbb{Q}$-linear span. 
We include the maximal possible number of the products \p{prodI} in the basis and complement them by a subset of \p{Iset}, $\{ I^{(w)}_{j_1},\ldots, I^{(w)}_{j_{L_w}} \}$ with $ L_w \leq M_w$. 
The linear span of the subset does not contain products of lower weights, i.e.\ it is irreducible.
Let us now relabel the iterated integrals by $1,\ldots, L_w$. 
Complementing the symbols to the complete solutions of the DEs by means of \p{solsplit} we would like to choose 
\begin{align}
{\cal I}^{(w)}_{1},\ldots,{\cal I}^{(w)}_{L_w} \label{cand}
\end{align}
as an irreducible set of iterated integrals at weight $w$. 
However, it could happen that not all weight-$w$ solutions of the DE are expressible in terms of \p{cand} and products of already classified lower-weight iterated integrals,
$\{ {\cal I}^{(w_1)}_i \}_{i=1}^{L_{w_1}}$ with  $w_1 < w$. 
We could encounter a solution ${\cal J}$ of the DE which is
expressible in the constructed basis at the symbol level, but it also contains ``beyond-the-symbol'' terms, 
\begin{align}
  {\cal J} = \sum_{k=1}^{L_w} b_k {\cal I}^{(w)}_{k} + 
      \underbrace{\sum_{n=1}^{L_1} \sum_{m=1}^{L_{w-1}} b_{n,m} {\cal I}^{(1)}_{n} {\cal I}^{(w-1)}_{m} + \ldots }_{\text{products of lower weights}} + 
        \sum_{w' = 1}^{w-1}\sum_a \kappa^{(w-w')}_a {\bar R}^{(w')}_a \,. \label{Jbad}
\end{align}
Here $b_k,\,b_{n,m},\ldots$ are rational numbers, and $\bar{R}^{(w')}_a$ is a $\mathbb{Q}$-linear span of weight-$w'$ with $w'<w$ iterated
integrals which have not been included in the weight-$w'$ basis of iterated integrals at the previous steps of the classification. 
Then to classify weight-$w$ iterated integrals we would need to reconsider the classification of all lower weights. 
Fortunately, this complication can be very easily resolved. 
We find that it is sufficient to only extend the set of weight-1 iterated integrals $\{ {\cal I}^{(1)}_{i} \}_{i = 1}^{L_1}$. 
These extra integrals do not appear in the weight-1 solutions $\vec f^{(1)}_{\tau,\sigma}$ (see \cref{solf1_g11}). 
But their powers and their products with $\{ {\cal I}^{(w_1)}_i \}_{i=1}^{L_{w_1}}$,  $w_1 <w$, take into account all ${\bar R}^{(w')}_a$ terms in \cref{Jbad}.

\begin{table}[t]
  \begin{center}
    \begin{tabular}{>{\centering\arraybackslash}m{6cm}cccc}
      \toprule 
      Weight & 1 & 2 & 3 & 4 \\
      \midrule 
      \# iterated integrals $\mod$ lower weights (symbols)  & 10|0 & 70|9 &  460|22 & 1185|277 \\
      \# irreducible iterated integrals & 10|0 & 15|9 & 90|21 & 316|156 \\
      \bottomrule
    \end{tabular}
  \end{center}
  \caption{Number of the parity $even|odd$ independent iterated-integral solutions of DEs \p{difeq} for all four topologies and all $5!$ orientations combined.}
  \label{tab_it_int}
\end{table}

As a result of this classification, we find that we need 24 weight-2 functions, 111 weight-3 functions and 472 weight-4 functions
to express all UT master integrals for all four topologies in all orientations (up to weight 4), see \cref{tab_it_int}. This counting of pentagon functions should be compared with the counting of integrable symbols summarized in table 1 from \cite{Chicherin:2017dob}. Predictably, integrating the DEs we find considerably fewer solutions than one obtains by imposing the second-entry restriction, inspired by the Steinmann relations \cite{Caron-Huot:2016owq,Dixon:2016nkn}, on the space of integrable symbols. In the rest of this section we elaborate the details of the pentagon function classification.

\subsection{Parity-even letters of the alphabet in the analyticity region}
\label{sec_letters_P0}
 
The iterated integrals \p{it_int_def} involve $\dd{\log}$ kernels $\dd\log(W_i)$, $i=1,\ldots,31$,
so while implementing integrations in the region $P_0$ \p{s12channel} we need to keep track of possible singularities of $\dd\log(W_i)$. 
Let us consider first the parity-even letters of the alphabet \p{pent_alph}. 
The parity-odd ones are discussed in \cref{1fodditint}.

Most of the parity-even letters have definitive sign within $P_0$,
\begin{equation}
\begin{tabular}{L|L|L}
W_1 = s_{12} > 0            &                               & W_{16} = -s_{13} >0          \\
W_2 = s_{23} < 0            &   W_{11} = s_{34}+s_{35}  >0  & W_{17} = -s_{24} >0          \\
W_3 = s_{34} > 0            &   W_{13} = -s_{35}-s_{45} <0  & W_{18} = -s_{35} <0          \\
W_4 = s_{45} > 0            &   W_{14} = s_{45} -s_{23} >0  & W_{19} = -s_{14} >0          \\
W_{5} = s_{15} <0           &   W_{15} = s_{15}-s_{34}  <0  & W_{20} = -s_{25} >0          \\
W_6 = s_{34}+s_{45} > 0     &                               &                              \\
W_8 = -s_{13} -s_{14} > 0   &                               & W_{24} = -s_{13} -s_{15} >0  \\
W_9 = s_{45} -s_{13} >0     &                               & W_{25} = s_{23} +s_{25} <0   \\
\end{tabular}
\label{fixsgn} 
\end{equation}
and $W_{31} = \ep_5 \equiv \ii \delta$ with $\delta > 0$ inside $P_0$. 
The corresponding $\dd\log(W_i)$ kernels are real-analytic inside $P_0$ and integration along a path $\gamma$, $\gamma \subset P_0$, is well defined. 
Missing in \cref{fixsgn} are the parity-even letters 
\begin{equation} \label{badletters}
  \begin{tabular}{LLL}
    W_7=v_4+v_5,                      &\qquad&   W_{21}= v_3+v_4  -v_1-v_2, \\
    W_{10}= v_2+v_3,                  &\qquad&   W_{22}= v_4+v_5 -v_2-v_3,  \\
    W_{12}= v_2-v_5,                  &\qquad&   W_{23}= v_1+v_5-v_3-v_4,   \\
  \end{tabular}
\end{equation}
which all vanish at the initial point $X_0$ \p{B}.
Since they are linear in the Mandelstam invariants, along any line segment $\gamma = [X_0;  X]$ parametrized by $0 \leq t \leq 1$ (see \cref{Xt}),
\begin{align}
W_k (X(t)) = b_k(X) \, t  
\;,\qquad k \in \alpha \coloneqq \{ 7,10,12,21,22,23 \} \,. \label{badWt}
\end{align}
where $b_k = b_k(X)$ is a constant along the path $\gamma$.
Thus, either $\dd\log W_k \equiv 0$ or $\dd\log W_k = \dd\log t$.
In the latter case, the $\dd\log$ kernel has a simple pole at $t = 0$, and we should verify that the corresponding integration in \cref{it_int_def} is well-defined.

\subsection{Weight-one solutions}

Weight-1 solutions of DEs \p{difeq} have a very simple form. According to \p{solDE}
\begin{align}
\vec f^{(1)}_{\tau,\sigma}(X) = \sum_{i = 1}^{31} a^{(i)}_{\tau,\sigma}  \vec f^{(0)}_{\tau} \left.[W_{i}]\right._{X_0}\!\!(X)  + \vec f^{(1)}_{\tau,\sigma}(X_0) \,. \label{solf1}
\end{align}
The previous equation involves only letters $\{ W_i \}_{i = 1}^5 \cup \{ W_i \}_{i=16}^{20}$\,, i.e.\ the vector $\vec f^{(0)}_{\tau}$ of rational numbers (see \p{f0sol}) is annihilated by the components of $\dd\tilde A_{\tau,\sigma}$ \p{difeq} corresponding to the remaining letters, 
\begin{align}
a^{(i)}_{\tau,\sigma}  \vec f^{(0)}_{\tau} = 0 \;,\qquad
i=6,\ldots,15,21,\ldots,31 \,. 
\end{align}
The one-fold integrals $\left.[W_i]\right._{X_0}\!\!(X)$ from \p{solf1} are calculated straightforwardly, see \p{it_int_def} and \p{B}. For example,
\begin{align}
\left.[W_1]\right._{X_0}\!\!(X) = \log(s_{12}) - \log(3) \;,\quad
\left.[W_2]\right._{X_0}\!\!(X) = \log(-s_{23}) \,. \label{it_int_1_exp}
\end{align}
The last term in $\p{solf1}$ is in the $\mathbb{Q}$-span of $\log(3),\ii\overset{\circ}{\pi}$ (see Tab.~\ref{tab_trans_const}). In fact, the transcendental constant $\log(3)$, which comes from the one-fold integrations and from weight-1 initial values, cancels out in \p{solf1}. 

The spurious transcendental constants related to the specific choice of $X_0$ \p{B} is one of the reasons why we prefer not to evaluate iterated integrals as in \p{it_int_1_exp}. Instead, we introduce a set of the following ten functions $\{ g^{(1)}_{1,i} \}_{i=1}^{10}$,
\begin{equation}\label{g11}
  \begin{aligned}
    g^{(1)}_{1,1}  &= \log(s_{12})\,,   &   g^{(1)}_{1,2} &= \log(-s_{23})\,,   &  g^{(1)}_{1,3}  &= \log(s_{34})\,, & \\
    g^{(1)}_{1,4}  &= \log(s_{45})\,,   &   g^{(1)}_{1,5} &= \log(-s_{15})\,,   &  g^{(1)}_{1,6}  &= \log(-s_{13})\,, &  \\
    g^{(1)}_{1,7}  &= \log(-s_{24})\,,  &   g^{(1)}_{1,8} &= \log(s_{35})\,,    &  g^{(1)}_{1,9}  &= \log(-s_{14})\,, & g^{(1)}_{1,10} = \log(-s_{25})\,. 
  \end{aligned}                                                                 
\end{equation}
They are well-defined in the analyticity region $P_0$ \p{s12channel}. 
The arguments of logarithms in \p{g11} are equal up to a sign to the letters $\{ W_i \}_{i = 1}^5 \cup \{ W_i \}_{i=16}^{20}$, and they are listed in the first and the third columns of \p{fixsgn}. 
We can assign even parity to $\{ g^{(1)}_{1,i} \}_{i=1}^{10}$. 
Then we represent functions \p{g11} as one-fold integrals with the initial point $X_0$, resolve the one-fold integrals in terms of functions \p{g11}, and substitute the former in \p{solf1}. 
In this way, we find weight-1 solutions $\vec{f}^{(1,\pm)}_{\tau,\sigma}$ \p{solf1} in the parity-even and parity-odd sectors
\begin{subequations}\label{solf1_g11}
  \begin{align}
  & \vec f^{(1,+)}_{\tau,\sigma}(X) = \sum_{i=1}^{10} \vec b_{\tau,\sigma,i} \, g^{(1)}_{1,i}(X) + \ii \overset{\circ}{\pi} \, \vec c_{\tau,\sigma}\,,  \\
  & \vec f^{(1,-)}_{\tau,\sigma}(X) = 0,
  \end{align}
\end{subequations}
where $\vec b$, $\vec c$ are vectors of rational numbers. Of course, at weight 1 the two approaches are completely equivalent, but at higher weights the second one is more practical.

\subsubsection{Extra weight-one functions}

As we have noted at the end of \cref{secStrategy}, we need to supplement functions \p{g11}, appearing in the weight-1 solutions \p{solf1_g11},
by some extra weight-1 functions that are needed to describe higher weight solutions of the DEs.

Let us start with the parity-even functions. We define ten functions which are logarithms of the remaining arguments from \p{fixsgn}
\begin{equation}\label{g12}
  \begin{aligned}
    g^{(1)}_{2,1} &= \log(s_{34}+s_{45}) \,,    &  g^{(1)}_{2,2}  &= \log(-s_{13}-s_{14}) \,, &  g^{(1)}_{2,3} &= \log(s_{45} -s_{13})  \,,  \\
   g^{(1)}_{2,4}  &= \log(s_{34}+s_{35}) \,,     &  g^{(1)}_{2,5} &= \log(s_{35}+s_{45}) \,,  &  g^{(1)}_{2,6} &= \log(s_{45} -s_{23}) \,, \\
   g^{(1)}_{2,7}  &= \log(s_{34} -s_{15}) \,,    &  g^{(1)}_{2,8} &= \log(-s_{13}-s_{15}) \,, &  g^{(1)}_{2,9} &= \log(-s_{23}-s_{25})  \,, \\
   g^{(1)}_{2,10} &= \log(W_{31}) = \log(\ii \delta)\,. & & \\ 
  \end{aligned}  
\end{equation}
They are well-defined everywhere inside $P_0$ \p{s12channel}. 
Let us note that we do not introduce logarithms of letters \p{badletters} which do not have definitive signs in $P_0$.
We assign positive parity to $\{ g^{(1)}_{2,i}\}_{i = 1}^{10}$. 

We will also need parity-odd weight-1 functions.
They are one-fold integrals \p{it_int_def} of the parity-odd letters $\{ W_{i}\}_{i=26}^{30}$ along a path $\gamma$ connecting $X_0$ and $X \in P_0$,
\begin{align}
& g^{(1)}_{3,k}(X) = \left.[W_{25+k}]\right._{X_0}\!\!(X)= \int_{\gamma} \dd \log W_{25+k} \,, \quad k = 1,\ldots,5  \label{g13}.
\end{align}
In the next section, we explain that they are well-defined and single-valued within $P_0$ \p{B} and provide explicit expressions \p{itintodd1}, \p{itintodd2} for them.
 
\subsubsection{One-fold iterated integrals of the parity odd letters} \label{1fodditint}

\begin{figure}[ht]
  \centering
  \begin{subfigure}[b]{0.4\textwidth}
    \includegraphics[width=0.8\textwidth]{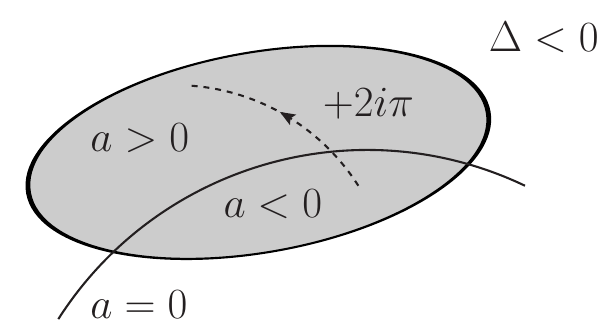}
  \end{subfigure}
  \begin{subfigure}[b]{0.4\textwidth}
    \includegraphics[width=0.8\textwidth]{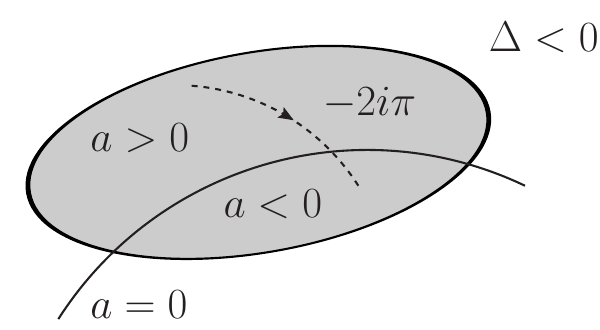}
  \end{subfigure}
  \caption{
    Definition of the single-valued logarithms of the parity odd letters in the physical region $\Delta < 0$ (shaded). 
    The discontinuity at $a_{k} = 0$ is added or subtracted depending on the direction of the branch-cut crossing.
  }
  \label{fig:dlogbranches}
\end{figure}

The parity odd-letters $\{ W_{i}\}_{i=26}^{30}$ have the following form  (see \cref{pent_alph,a_def}),
\begin{align}
W_{25+k}(X) = \frac{a_{k} - \ii \delta }{a_{k} + \ii \delta } = \exp(\ii \varphi_k(X)) \;,\qquad k = 1 ,\ldots,5 \label{Wodd_phase}
\end{align}
with $\ep_5 = \ii \delta$ and $\delta > 0$ and real $a_k$ inside $P_0$ \p{B}, so they are pure phases \p{phase}. Then integrals \p{g13} evaluate to
\begin{align}
g_{3,k}^{(1)}(X) = \ii\varphi_k(X) - \ii\varphi_k(X_0) 
\end{align}
provided the phases $\varphi_k$ do not have discontinuities inside $P_0$. At the initial point $X_0$ \p{B} we have $\delta = \sqrt{3}$ and
\begin{align}
a_{1}(X_0) = a_{2}(X_0) = a_{3}(X_0) = 3 \,,\qquad
a_{4}(X_0) = a_{5}(X_0) = -1 \,, 
\end{align}
then choosing the principal branch\footnote{We define the principal branch of the logarithm such that $\log(e^{\ii\varphi}) = \ii\varphi$ for $-\pi < \varphi \leq \pi$.} of the logarithm we find
\begin{align}
& \varphi_{1}(X_0) = \varphi_{2}(X_0) = \varphi_{3}(X_0) = - \frac{ \pi}{3}\,, \qquad 
\varphi_{4}(X_0) = \varphi_{5}(X_0) =  \frac{2\pi}{3} \,.
\end{align} 

Let us now define continuous phases $\varphi_k$ \p{Wodd_phase} inside $P_0$. The phase $\varphi_k$ never takes values $0,\pm 2\pi, \pm 4\pi, \ldots$. Indeed,
\begin{align}
{\rm Re}\, W_{25+k} = \frac{a_{k}^2 - \delta^2}{a_{k}^2 + \delta^2} \, , \qquad 
{\rm Im}\, W_{25+k} = - \frac{2 a_{k}  \delta}{a_{k}^2 + \delta^2} \,. \label{ReImOddW}
\end{align}
with $\delta > 0$ in the analyticity region $P_0$ \p{s12channel}, and
${\rm Im}\,W_{25+k} = 0$ implies ${\rm Re}\, W_{25+k} = -1$. 
Thus, $-2\pi<\varphi_{k} < 0$ at $k= 1,2,3$, and $0< \varphi_{k} < 2\pi$ at $k=4,5$. 
We need to match them with the principal branch of the logarithm. 
We cross the branch cut and go to another Riemann sheet of the logarithm only at $a_{k} = 0$, see \p{ReImOddW}. 
If we go from the region $a_{k} >0$ to the region $a_{k} < 0$, then we decrease the phase $\varphi_k$, and we should add $-2 \ii \pi$ to the principal value of the logarithm.  
If we go from the region $a_{k} < 0$ to the region $a_{k} > 0$, then we increase the phase $\varphi_k$, so we should add $+ 2 \ii \pi$ to the principal value of the logarithm.
We illustrate this in \cref{fig:dlogbranches}

Thus we obtain the following following continuous and real-analytic expressions for integrals \p{g13} inside $P_0$ \p{s12channel}:
\begin{align}
g^{(1)}_{3,k}(X) \, =  \, \theta(a_{k}) \, \log W_{25+k} +
\theta(-a_{k}) \, (\log W_{25+k} - 2\ii \pi)  -\ii \pi \, \delta_{0 a_{k}} + \frac{\ii \pi}{3} \label{itintodd1}
\end{align}
for $k=1,2,3$, and
\begin{align}
g^{(1)}_{3,k}(X) \, =  \, \theta(-a_{k}) \, \log W_{25+k} +
\theta(a_{k}) \, (\log W_{25+k} + 2\ii \pi) + \ii \pi \, \delta_{0 a_{k}} - \frac{2\ii\pi}{3} \label{itintodd2}
\end{align}
for $k=4,5$. Here $\delta_{ij}$ is the Kronecker delta, and the function $\theta(x)$ is defined as
\begin{equation}
  \theta(x) = 
  \begin{cases}
    1, & x>0,\\  
    0, & x\leq 0,\\  
  \end{cases}
\end{equation}

\subsection{Weight-two solutions}

According to \p{solDE}, weight-2 solutions of DEs \p{difeq} can be represented as
\begin{multline}
\vec f^{(2)}_{\tau,\sigma}(X) = \\
\sum_{i,j = 1}^{31} a^{(i)}_{\tau,\sigma} a^{(j)}_{\tau,\sigma}  \vec f^{(0)}_{\tau} \left.[W_{i},W_j]\right._{X_0}\!\!(X) + \sum_{i = 1}^{31} a^{(i)}_{\tau,\sigma} \vec f^{(1)}_{\tau,\sigma}(X_0) \left.[W_{i}]\right._{X_0}\!\!(X)   + \vec f^{(2)}_{\tau,\sigma}(X_0) \,.\label{solf2}
\end{multline}
The first term in the previous equation corresponds to the symbol of the solution. 
Instead of evaluating one-fold and two-fold iterated integrals from \p{solf2}, and then looking for cancellation of the spurious transcendental constants among the three terms in \p{solf2},
we prefer to start with a set of 15 parity-even and 9 parity-odd weight-2 functions and to express \p{solf2} in terms of them.

\subsubsection{Parity-even functions}
\label{secg21}

We introduce 15 weight-2 parity-even pentagon functions $\{ g^{(2)}_{1,i} \}_{i = 1}^{15}$,
\begin{equation} \label{g21}
  \begin{aligned}
    &\begin{aligned}
      g^{(2)}_{1,1}  &= {\rm Li}_2\left(1-\frac{s_{34}}{s_{12}}\right), &
      g^{(2)}_{1,2}  &= {\rm Li}_2\left(1-\frac{s_{45}}{s_{12}}\right), & 
      g^{(2)}_{1,4}  &= {\rm Li}_2\left(1-\frac{s_{15}}{s_{23}}\right),   \\
      g^{(2)}_{1,7}  &= {\rm Li}_2\left(1-\frac{s_{24}}{s_{15}}\right), &
      g^{(2)}_{1,8}  &= {\rm Li}_2\left(1-\frac{s_{24}}{s_{13}}\right), &
      g^{(2)}_{1,9}  &= {\rm Li}_2\left(1-\frac{s_{35}}{s_{12}}\right),  \\
      g^{(2)}_{1,11} &= {\rm Li}_2\left(1-\frac{s_{14}}{s_{23}}\right), &
      g^{(2)}_{1,14} &= {\rm Li}_2\left(1-\frac{s_{25}}{s_{13}}\right), &
      g^{(2)}_{1,15} &= {\rm Li}_2\left(1-\frac{s_{25}}{s_{14}}\right), &
    \end{aligned}\\
    &\begin{aligned}
      &g^{(2)}_{1,3} = -{\rm Li}_2\left(\frac{s_{45}}{s_{23}}\right) - \log\left(1-\frac{s_{45}}{s_{23}} \right) \log\left( -\frac{s_{45}}{s_{23}}\right) , \\ 
      & g^{(2)}_{1,5} = -{\rm Li}_2\left(\frac{s_{15}}{s_{34}}\right) - \log\left(1-\frac{s_{15}}{s_{34}} \right) \log\left( -\frac{s_{15}}{s_{34}}\right) , \\ 
      & g^{(2)}_{1,6} = -{\rm Li}_2\left(\frac{s_{13}}{s_{45}}\right) - \log\left(1-\frac{s_{13}}{s_{45}} \right) \log\left( -\frac{s_{13}}{s_{45}}\right) , \\ 
      & g^{(2)}_{1,10} = -{\rm Li}_2\left(\frac{s_{35}}{s_{24}}\right) - \log\left(1-\frac{s_{35}}{s_{24}} \right) \log\left( -\frac{s_{35}}{s_{24}}\right) , \\ 
      & g^{(2)}_{1,12} = -{\rm Li}_2\left(\frac{s_{14}}{s_{35}}\right) - \log\left(1-\frac{s_{14}}{s_{35}} \right) \log\left( -\frac{s_{14}}{s_{35}}\right) , \\
      & g^{(2)}_{1,13} = -{\rm Li}_2\left(\frac{s_{25}}{s_{34}}\right) - \log\left(1-\frac{s_{25}}{s_{34}} \right) \log\left( -\frac{s_{25}}{s_{34}}\right) \,. 
    \end{aligned}
  \end{aligned}
\end{equation}
They are well-defined inside $P_0$ \p{s12channel}. Indeed, the arguments of $\log$ are positive, and the arguments of ${\rm Li}_2$ are less than $1$, so no branch cuts are crossed.

\subsubsection{Parity-odd functions}
\label{secg22}

In order to describe the weight-2 parity-odd pentagon functions, it is helpful to introduce the following combination
\begin{align}
\psi(a,b) \coloneqq  2\ii ( {\rm Cl}_2(a) + {\rm Cl}_2(b) + {\rm Cl}_2(-a-b))
\end{align}
of the order-two Clausen functions ${\rm Cl}_2$. 
The latter are defined by the dilogarithm evaluated on the unit circle,
\begin{equation}
 {\rm Cl}_2(\varphi) = \frac{1}{2\ii} \qty({\rm Li_2}(e^{\ii\varphi}) - {\rm Li_2}(e^{-\ii\varphi})), \quad {\rm Cl}_2(\varphi+2\pi) = {\rm Cl}_2(\varphi),
\end{equation}
and thus it is single-valued on the circle.
The parity-odd letters inside the analyticity region $P_0$ are pure phases $\{ e^{\ii \varphi_k(X)} \}_{k=1}^5$, see \p{phase}. 
We then introduce nine parity-odd functions $\{ g^{(2)}_{2,i}\}_{i = 1}^9$,
\begin{equation}\label{g22}
  \begin{aligned}
  g^{(2)}_{2,1}   &= \psi(-\varphi_2, -\varphi_3) \,, & g^{(2)}_{2,2}     &= \psi(-\varphi_3, -\varphi_4)  \,, \\
  g^{(2)}_{2,3}   &= \psi(-\varphi_2 - \varphi_3,\varphi_3 + \varphi_4-\varphi_1) \,,\qquad & g^{(2)}_{2,4}     &=   \psi( -\varphi_1 , - \varphi_5) \,, \\
  g^{(2)}_{2,5}   &= \psi(\varphi_1 + \varphi_2,\varphi_3 - \varphi_1 - \varphi_5) \,, & g^{(2)}_{2,6}     &= \psi(-\varphi_4,-\varphi_5) \,, \\
  g^{(2)}_{2,7}   &=  \psi(\varphi_4 + \varphi_5,\varphi_3 - \varphi_1 - \varphi_5) \,, & g^{(2)}_{2,8}     &= \psi(- \varphi_1 - \varphi_5,\varphi_1 + \varphi_2 - \varphi_4) \,, \\
  g^{(2)}_{2,9}   &= \psi(-\varphi_3 - \varphi_4,\varphi_2 + \varphi_3 - \varphi_5) \,.  & 
  \end{aligned}
\end{equation}
They belong to the orbit of 
\begin{equation}
 \psi(\varphi_1,\varphi_2) 
\end{equation}
under the action of the permutation group ${\cal S}_5$ on the external momenta.
In fact, the orbit consists of ten functions that are not trivially equivalent.
However, a nontrivial combination of the ten $\psi$ functions vanishes \cite{Chicherin:2017dob}, 
\begin{align}
\sum_{\sigma \in {\cal S}_5} (-1)^{\sgn(\sigma)} \psi(\sigma(\varphi_1),\sigma(\varphi_2)) = 0 \,
\end{align}
or more explicitly
\begin{align}
\sum_{i=1}^5 \left[ - \psi(\varphi_{i},\varphi_{i+1}) + \psi(\varphi_{i} + \varphi_{i+1} , \varphi_{i+2} - \varphi_{i-1} - \varphi_{i} ) \right] = 0 \,.
\end{align}
This allows us to choose nine linear independent functions \eqref{g22} from the orbit.

\subsubsection{All master UT integrals at weight two}
\label{sec_w2_sol}

Now we have all necessary ingredients to express weight-2 solutions $\vec f_{\tau,\sigma}^{(2)}$. 
We transform 15 parity-even \p{g21} and 9 parity-odd \p{g22} functions into the iterated-integral representation which involves one-fold $[W_i]_{X_0}(X)$
and two-fold $[W_i,W_j]_{X_0}(X)$ iterated integrations. 
We also transform all weight-1 functions in \cref{g11,g12,g13} into the iterated integral representation, i.e.\ one-fold integrations. 
Then we resolve the iterated integrals $[W_i]_{X_0}$ and $[W_i,W_j]_{X_0}$ for the weight-1 and weight-2 functions,
and, by substituting these relations into \p{solf2}, we express all weight-2 solutions in terms of the functions $g$. 

The parity-even and parity-odd solutions, defined in \cref{sec_parity}, involve different subsets of functions. 
We find that the parity-even solutions have the following form
\begin{subequations}
  \begin{align}
    \vec f^{(2,+)}_{\tau,\sigma}  = \sum_{i = 1}^{15} \vec b_{\tau,\sigma,i}\, g^{(2)}_{1,i} + \sum_{i,j = 1}^{10} \vec b_{\tau,\sigma,i,j} \, g^{(1)}_{1,i} g^{(1)}_{1,j}  + \ii\overset{\circ}{\pi} \sum_{a=1}^2\sum_{i} \vec c_{\tau,\sigma,a,i} \, g^{(1)}_{a,i}  + \vec d_{\tau,\sigma} \overset{\circ}{\pi}^2 
    \label{solf2_g_even}
  \end{align}
  and the parity odd ones
  \begin{align}
    \vec f^{(2,-)}_{\tau,\sigma} = \sum_{i = 1}^{9} \vec b_{\tau,\sigma,i} \,  g^{(2)}_{2,i} + \ii\overset{\circ}{\pi} \sum_{i=1}^{5} \vec c_{\tau,\sigma,i} \, g^{(1)}_{3,i}  + \vec d_{\tau,\sigma} \pi \overset{\circ}{\pi}
    \label{solf2_g_odd}
  \end{align}
\end{subequations}
where $\vec b$, $\vec c$, and $\vec d$ are vectors of rational numbers. 
These expressions respect $\mathbb{Z}_{\geq 0} \times \mathbb{Z}_2$-grading. 
As we can see upon identifying two copies of $\ii\pi$ the only transcendental constants in the solution are $\ii\pi$ and $\pi^2$.
In fact, not all extra weight-1 functions \p{g12}, \p{g13} are present. 
The weight-2 solutions contain only parity even $g^{(1)}_{2,k}$ with $k = 2,3,6,7,8,9$ and parity odd $g^{(1)}_{3,k}$ with $k = 1,3,4,5$.

As we can see, the weight-1 functions in \cref{g12,g13} that are not needed to express the weight-1 solutions \p{solf1_g11}
start appearing in the weight-2 solutions \cref{solf2_g_even,solf2_g_odd} in the ``beyond-the-symbol'' part.
This phenomenon illustrates the statements from \cref{secStrategy}.

\subsection{Weight-three solutions}

\subsubsection{Weight-three pentagon functions}
\label{subsec_w3_pentagon_func}

We follow the classification procedure from \cref{secStrategy} and find 90 parity-even and 21 parity-odd weight-3 irreducible iterated integrals. 
Let us recall that they have the form of \cref{solsplit} with $w = 3$. 
At weights one and two we started with the nice choices of logarithmic and dilogarithmic pentagon functions and expressed all iterated integrals in terms of these functions. 
It was argued in \cite{Caron-Huot:2014lda} that at higher weights explicit representations of iterated integrals in terms of MPLs are not always beneficial,
especially for the purpose of their efficient numerical evaluation. We find that this also applies to the iterated integrals studied in this paper (see \cref{sec:alternative_representation}).
We then choose a set $\{ g^{(3)}_i \}_{i=1}^{111}$ of irreducible iterated integrals as our weight-3 pentagon functions. 
Taken together with already classified iterated integrals at weights one and two, they are sufficient to express any weight-3 solution $\vec f^{(3)}_{\tau,\sigma}$ in \cref{f_it_int_gen}. 
The functions $\{ g^{(3)}_i \}_{i=1}^{111}$ take the form of a one fold integral along a path $\gamma$ connecting $X_0$ \p{B} and $X \in P_0$,
\begin{align}
g^{(3)}_i(X) = \sum_{j=1}^{31} \int_\gamma \dd\log W_{j}(X') \, h^{(2,\pm)}_{i,j}(X') \;,\qquad i=1,\ldots,111 , \label{w3onefold}
\end{align}
where $h^{(2,\pm)}_{i,j}$  are weight-2 polynomials of definite parity,
\begin{align}
& h^{(2,+)} \in \mathbb{Q}\left[ \ii\overset{\circ}{\pi}, \{ g^{(2)}_{1,i} \}_{i=1}^{15} , \{ g^{(1)}_{1,i} \}_{i=1}^{10} , \{ g^{(1)}_{2,i} \}_{i=2,3,6,7,8,9} \right]_{(2,+)} , \notag\\
& h^{(2,-)} \in \mathbb{Q}\left[ \ii\pi, \ii\overset{\circ}{\pi}, \{ g^{(2)}_{2,i} \}_{i=1}^{9} , \{ g^{(1)}_{3,i} \}_{i=1,3,4,5} \right]_{(2,-)} . \label{h2}
\end{align}
In fact, the parity-even and parity-odd $h^{(2)}$-functions have exactly the same form as \p{solf2_g_even} and \p{solf2_g_odd} weight-2 solutions, respectively. 
The iterated integrals $\{ g^{(3)}_i \}_{i=1}^{111}$ by construction have definite parity induced by the $\mathbb{Z}_2$-grading, which was introduced in \cref{sec_parity}.
The parity-even weight-3 pentagon functions have the form 
\begin{subequations}
  \begin{align} \label{eq:g33p}
    g^{(3,+)}_i  = 
    \sum_{j=1}^{25} \int_\gamma \dd\log W_{j} \cdot h^{(2,+)}_{i,j}  + 
    \sum_{j=26}^{30} \int_\gamma \dd\log W_{j} \cdot h^{(2,-)}_{i,j}
  \end{align}
  and the parity-odd ones
  \begin{align} \label{eq:g33m}
    g^{(3,-)}_i = 
    \sum_{\substack{j \in \{1 ,\ldots,5,\\ 16,\ldots,20,31 \} }} \int_\gamma \dd\log W_{j} \cdot h^{(2,-)}_{i,j} + 
    \sum_{j=26}^{30} \int_\gamma \dd\log W_{j} \cdot h^{(2,+)}_{i,j}  \,.
  \end{align}
\end{subequations}

We must verify that the integrations in \cref{w3onefold} are well-defined. 
As in \cref{Xt}, we choose the path $\gamma$ as the line segment $\gamma = [X_0;X]$. 
The functions  $h^{(2)}_{i,j}$ are well-defined on $\gamma$, since they are polynomials of weight-1 and weight-2 pentagon functions.
The only possible source of potential problems is $\dd\log W_k$
with $k \in \alpha \coloneqq \{ 7,10,12,21,22,23\}$, i.e.\ $\dd{\log}$-forms of the letters that do not have a definite sign in the analyticity region $P_0$ (see \cref{badletters}). 
As we showed in \cref{badWt}, on the path $X(t)$ either they are $\dd\log(t)$ with a pole at $t= 0$ or they are identically zero. 
Fortunately, the pole is compensated by the accompanying
$h^{(2)}_{i,k}$ which vanishes at $t = 0$. 
Indeed, inspecting $h^{(2)}_{i,k}$ for $ k \in \alpha$ we find that they
involve very simple pentagon functions: $\{ g^{(1)}_{1,i} \}_{i = 1}^{10}$ (\cref{g11}) and $g^{(2)}_{1,i}$ for $i = 4,7,8,11,14,15$ (\cref{g21}). 
Resolving $W_k = 0$ \p{pent_alph} as a constraint on $\{ v_i\}_{i=1}^{5}$ and evaluating $h^{(2)}_{i,k}$ on this subspace we find that it vanishes, 
\begin{align} \label{eq:vanishing-dlog-w3}
\left. h^{(2)}_{i,k} \right|_{W_k = 0} = 0 \,,\qquad k \in \alpha \,.  
\end{align} 
Thus the integrations in the definition of the weight-3 pentagon functions \p{w3onefold} are well-defined for any point of the analyticity domain $P_0$ (see \cref{s12channel}).

\subsubsection{All master UT integrals at weight three}

We are now ready to express the weight-3 solutions \p{solDE} of DEs \p{difeq} in terms of the classified pentagon functions
of weights one (\cref{g11,g12,g13}), two (\cref{g21,g22}), and three (\cref{w3onefold}), and transcendental constants ${\cal S}_0$ from \cref{tab_trans_const}. 
All steps of our construction respect the grading $\mathbb{Z}_{\geq 0} \times \mathbb{Z}_2$. 
The parity-even solutions take the following form
\begin{subequations}
  \begin{multline} \label{solw3_even}
    \vec f^{(3,+)}_{\tau,\sigma} \in \mathbb{Q} 
    \biggl[ g^{(3,+)} , \{ g^{(2)}_{1,i} \}_{i=1}^{15} , \{ g^{(1)}_{1,i} \}_{i=1}^{10} , \{ g^{(1)}_{2,i} \}_{i=2,3,6,7,8,9} , \\
    \ii\overset{\circ}{\pi},{\cal S}_{0}^{(1,+)} , {\cal S}_{0}^{(2,+)},{\cal S}_{0}^{(3,+)} \biggr]_{(3,+)} 
  \end{multline}
  and the parity odd ones
  \begin{multline}
    \vec f^{(3,-)}_{\tau,\sigma} \in \mathbb{Q}
    \biggl[ g^{(3,-)},  \{ g^{(2)}_{2,i} \}_{i=1}^{9} , \{ g^{(1)}_{1,i} \}_{i=1}^{10}  ,  g^{(1)}_{2,10} , \{ g^{(1)}_{3,i} \}_{i=1,3,4,5} ,  \\
      \ii\pi,\ii\overset{\circ}{\pi}, \log(3),  {\cal S}_{0}^{(2,-)},
    \ii \Im\Li_3\left(\frac{\ii}{\sqrt{3}}\right)  \biggr]_{(3,-)} \,. \label{solw3_odd}
  \end{multline}
\end{subequations}
Explicit expressions for the UT master integrals are provided in the ancillary files (see \cref{sec:math_package}).
Let us note that one new extra weight-1 function $g^{(1)}_{2,10}$ appears in the weight-3 solution as compared to the weight-2 solution in \cref{solf2_g_even,solf2_g_odd}.
Of course, the extra weight-1 functions \p{g12} and \p{g13} appear only in the ``beyond-the-symbol'' part of the solution.

\subsection{Weight-four solutions}

\subsubsection{Weight-four pentagon functions}
 
At weight 4 the classification procedure from \cref{secStrategy} results in 316 parity-even and 156 parity-odd irreducible iterated integrals  having the form of \cref{solsplit} with $w = 4$. 
As at weight 3, we choose a set of 472 irreducible iterated integrals $\{ g^{(4)}_{i}\}_{i = 1}^{472}$ as our weight 4 pentagon functions.
To bring them into a more explicit form, we use the definition in \cref{it_int_def} and perform the two innermost integrations.
The functions  $g^{(4)}_i$ are then expressed as two-fold iterated integrals over functions $g^{(1)}_{i,j}$ and $g^{(2)}_{i,j}$ introduced above,
\begin{multline}
  g^{(4)}_i(X) = \\
    \sum_{j,k=1}^{31} \int_{\gamma} \dd\log W_{j}(X') \int_{\gamma} \dd\log W_{k}(X'') \, h^{(2)}_{i,j,k}(X'') +  
      \sum_{j=1}^{31}\kappa^{(3)}_{j} \int_{\gamma} \dd\log W_{j}(X') \,. \label{w4onefold'}
\end{multline}
where the weight-2 functions $h^{(2)}_{i,j,k}$ have definite parity, and they are of the same form as in \cref{h2}. 
The transcendental constants $\kappa^{(3)} \in \mathbb{Q}[{\cal S}_0]_{(3,\pm)}$ are given in \cref{tab_trans_const}. 
\Cref{w4onefold'} respects $\mathbb{Z}_{\geq 0} \times \mathbb{Z}_2$-grading, i.e.\ the transcendental weight and parity counting. 
In order to render the weight-4 pentagon functions \p{w4onefold'} to a form better adapted for numerical evaluations,
we rewrite them as one-fold integrals in the next section.

\subsubsection{One-fold integral representation of weight-four pentagon functions}
\label{sec:w4onefold}

We apply the technique from \cite{Caron-Huot:2014lda} to rewrite the two-fold iterated integrals in \p{w4onefold'} into one-fold integrals. 
We introduce parametrization \p{Xt} of the path $\gamma$ which we choose as the line segment, $\gamma = [X_0;X]$, such that $X(1) = X$ and $X(0)=X_0$, and interchange the order of integrations  
\begin{align}
& \int^1_0 \dd t \,\pa_t \log(W_j(t)) \int^t_0 \dd u \, \pa_u \log(W_k(u)) \, h^{(2)}_{i,j,k}(X(u)) \notag \\
& = \int^1_0 \dd u \int^1_u \dd t \,  \pa_t \log(W_j(t)) \cdot  \pa_u \log(W_k(u)) \, h^{(2)}_{i,j,k}(X(u)) \, \label{swapint}
\end{align}
where $W_i(t) \coloneqq W_i(X(t))$.
Thus, one of the integrations in \cref{swapint} becomes trivial, and naively we obtain 
\begin{align}
\int^1_u \dd t \,  \pa_t \log(W_j(t)) = \log(W_j(u=1)) - \log( W_j(u)) \,. \label{intdlog}
\end{align}
However, the right-hand-side of \p{intdlog} should be well-defined for $0 \leq  u \leq 1$. 
It is straightforward to express it in terms of the weight-1 pentagon functions (\cref{g11,g12,g13}). 
Indeed, for the parity-odd letters we use \cref{itintodd1,itintodd2},
\begin{align}
\int^1_u \dd t \,  \pa_t \log(W_j(t)) = g^{(1)}_{3,j-25}(X(1)) - g^{(1)}_{3,j-25}(X(u)) \,, \qquad j = 26 ,\ldots,30 \,. 
\end{align}
For the parity-even letters that are linear in the Mandelstam invariants 
\begin{multline} \label{eq:intdlogsp}
  \qquad\int^1_u \dd t \,  \pa_t \log(W_j(t)) = \log( s_j \cdot W_j(1)) - \log( s_j \cdot W_j(u)) \, , \\
  s_j  \coloneqq \sgn(W_j(1)) \,, \quad j = 1,\ldots, 25 \,. 
\end{multline}
Let us note that here we have to consider letters \p{badletters}, i.e.\ $j \in \alpha \coloneqq \{ 7,10,12,21,22,23 \} $, which do not have definitive sign in $P_0$. 
Still, they do have definitive sign on the line segment $\gamma$, see \p{badWt}, and they vanish at $u = 0$. 
Unlike other letters, which do not vanish inside $P_0$, they introduce logarithmic singularity at $u = 0$, and thus they are not among pentagon functions \p{g11} and \p{g12}. 
Finally, for the remaining parity-even letter $W_{31}$,
\begin{align}
  \int^1_u \dd t \,  \pa_t \log(W_{31}(t)) = \log( \delta(1)) - \log( \delta(u)) \, . \label{intdlogw31}
\end{align}
In this away, we arrive at the one-fold integral representation of the weight-4 pentagon functions \p{w4onefold'},
\begin{align}
g^{(4)}_i(X) = \sum_{k=1}^{31}\int^1_0 \dd u \,\pa_u 
    \log(W_k(u)) \left[ \int^1_u \dd t \,  \pa_t \log(W_j(t)) \cdot  
        h^{(2)}_{i,j,k}(X(u)) + \kappa^{(3)}_k \right]\,. \label{w4onefold2}
\end{align}
We need to verify that integrations in \p{w4onefold2} are well-defined. 
The analysis is similar to the one from \cref{subsec_w3_pentagon_func}. 
The weight-2 functions $h^{(2)}$ are polynomials in the pentagon functions as in \p{h2}, and they are real-analytic. 
A potential source of pole singularities in the integrand is $\pa_u \log(W_k(u))$ at $k \in \alpha$. 
We find that the pole is suppressed since
\begin{align} \label{eq:vanishing-dlog-w4}
\left. h^{(2)}_{i,j,k} \right|_{W_k = 0} = 0 \;,\quad \kappa^{(3)}_{k} = 0 \;,\qquad \text{for} \quad k\in \alpha \,.
\end{align}

Summarizing, we find that the integrations in \cref{w4onefold2} are well-defined. 
All terms of the integrands are real-analytic at $0 \leq u \leq 1$, except
for the terms with $j \in \alpha$, which are real-analytic at $0 < u \leq 1$.
The only singularity of the integrands is the logarithmic singularity $\log(u)$ in the terms with $j \in \alpha$.
This singularity is integrable.
Nevertheless, some care should be taken in an algorithm for numerical evaluations, see \cref{sec:cpplib} for details.

\subsubsection{All master UT integrals at weight four}

We express all weight-4 solutions \p{solDE} of the DEs as homogeneous polynomials of definite parity in the pentagon functions 
of weight one \p{g11}, \p{g12}, \p{g13}, weight two \p{g21}, \p{g22}, weight three \p{w3onefold}, and weight four \p{w4onefold'}, 
and in transcendental constants from \cref{tab_trans_const}, and we find that the parity-even solutions have the following form
\begin{subequations}
  \begin{align} \label{solw4_even}
    \vec f^{(4,+)}_{\tau,\sigma} \in \mathbb{Q}\biggl[ & g^{(4,+)}, g^{(3,+)} , \{ g^{(2)}_{1,i} \}_{i=1}^{15} , \{ g^{(2)}_{2,i} \}_{i=1}^{9}  , \{ g^{(1)}_{1,i} \}_{i=1}^{10} , \{ g^{(1)}_{2,i} \}_{i=2,3,6,7,8,9}, \{ g^{(1)}_{3,i} \}_{i=1,3,4,5} ,\notag\\
                                                       & {\cal S}^{(4,+)}_{0} , {\cal S}^{(3,+)}_{0} , {\cal S}^{(2,\pm)}_{0} , {\cal S}^{(1,\pm)}_{0} \biggr]_{(4,+)} 
    \end{align}
    and the parity odd ones
    \begin{align}
      \vec f^{(4,-)}_{\tau,\sigma} \in \mathbb{Q}\biggl[ & g^{(4,-)}, g^{(3,-)},  \{ g^{(2,)}_{2,i} \}_{i=1}^{9} , \{ g^{(1)}_{1,i} \}_{i=1}^{10}  ,  g^{(1)}_{2,10} , \{ g^{(1)}_{3,i} \}_{i=1,3,4,5} ,\notag\\
                                                         & {\cal S}_{0}^{(4,-)},{\cal S}_{0}^{(3,-)}, \zeta_3, {\cal S}_{0}^{(2,-)}, {\cal S}_{0}^{(1,\pm)} \biggr]_{(4,-)} \,. \label{solw4_odd}
    \end{align}
\end{subequations}
Explicit expressions are provided in the ancillary files (see \cref{sec:math_package}).
It is worth noting that not all of the allowed by the $\mathbb{Z}_{\geq 0} \times \mathbb{Z}_2$-grading terms are present in the solutions. 

Thus, we have classified all pentagon functions up to transcendental weight four,
and we identified the minimal generating set in the pentagon function space. 
All constructed pentagon functions are well-defined within the physical region $P_0$ \p{s12channel}. 
We also provided $\ep$-expansion of all two-loop UT master integrals that describe the massless five-particle scattering in terms of the pentagon functions.

\subsection{Alternative representation of the pentagon functions}
\label{sec:alternative_representation}

We provided expressions for the pentagon functions in terms of the familiar polylogarithmic functions only at weights one and two, i.e.\ logarithms and dilogarithms, respectively. We preferred to express the pentagon functions of higher weights as one-fold integrations, see \p{w3onefold} and \p{w4onefold2}. 
This approach provides a convenient setup for numerical evaluations of the pentagon functions,
and thus it is completely sufficient for all imaginable phenomenological applications of the pentagon functions. 
We implemented this approach in a public \texttt{C++} library and a public \texttt{Mathematica} package, which we describe in \cref{sec:math_package,sec:cpplib}.

Nevertheless, one could ask a question how to express weight three and four pentagon functions in terms of polylogarithmic functions. 
We found expressions for all $90|21$ weight-3 pentagon functions \p{w3onefold} in terms of logarithms, dilogarithms and trilogarithms with arguments built from the letters of the pentagon alphabet. 
Using this weight-3 result it is straightforward to obtain and alternative integral representation for all $315|156$ weight-4 pentagon functions \p{w4onefold'}. 
Indeed, we explicitly implement all inner three-fold iterated integrations and obtain 
\begin{align}
g^{(4)}_i (X) = \sum_{j=1}^{31} \int_\gamma \dd \log W_j(X') \, h^{(3)}_{i,j}(X')
\end{align} 
where $h^{(3)}$ are  weight-3 polylogarithmic functions, which involve logarithms, dilogarithms and trilogarithms. 
Thus, we have at hand two alternative ways to evaluate weight-3 and weight-4 pentagon functions --- the one extensively described in the previous subsections and the one briefly outlined in this subsection. 
We stick to the first approach and use our private \texttt{Mathematica} implementation of the second approach as a highly-nontrivial test for the public library and the public package.

\subsection{Master integrals in arbitrary channel}
\label{secAnyChannel}
All previous considerations were restricted to the subset $P_0$ \p{s12channel} of the physical region,
the $s_{12}$-scattering channel.
We classified pentagon functions up to weight four, which are well-defined inside $P_0$, and we provided $\ep$-expansion of all UT master integrals in all orientations in terms of the pentagon functions. 
Thus, we are able to evaluate all master integrals within $P_0$. 
In this section we demonstrate that having at hand results for all $5!$ orientations of the master integrals is equivalent to knowing them in any physical region.

Let $X$ be a kinematic point in an arbitrary scattering channel of the physical region.
We can always find an element $\tilde{\sigma}$ in $\mathcal{S}_5$ which maps the point $X$ into a
point $\tilde{X}$ from the $s_{12}$-scattering channel,
\begin{equation}
  \tilde{X} = \tilde{\sigma} X  ~\in~ P_{0}.
\end{equation}
The previous equation implies that $X$ belongs to the scattering channel $\bar \sigma_1 \bar\sigma_2 \to  \bar\sigma_3 \bar\sigma_4 \bar\sigma_5$ where $\bar\sigma \coloneqq (\tilde\sigma)^{-1}$
and the following inequalities which specify the scattering channel hold
\begin{align}
& s_{\bar\sigma_1 \bar\sigma_2} ,\, s_{\bar\sigma_3 \bar\sigma_4} ,\,  
s_{\bar\sigma_3 \bar\sigma_5},\,
s_{\bar\sigma_4 \bar\sigma_5} > 0 \,,\notag\\
X: \qquad  & s_{\bar\sigma_1 \bar\sigma_3} ,\,
s_{\bar\sigma_1 \bar\sigma_4} ,\,
s_{\bar\sigma_1 \bar\sigma_5} ,\,
s_{\bar\sigma_2 \bar\sigma_3} ,\,
s_{\bar\sigma_2 \bar\sigma_4} ,\,
s_{\bar\sigma_2 \bar\sigma_4} < 0\,, \notag\\
& {\rm sgn}(\bar\sigma) \, \bar\sigma(\delta)  > 0 \,.
\end{align}
  
Then we can use \cref{fperm,eq:perm-composition} to evaluate master integrals $\vec{f}_{\tau,\sigma}$  (in arbitrary orientation $\sigma$) at the point $X$ from the arbitrary scattering channel
as follows,
\begin{equation} \label{eq:mi-any-channel}
  \vec{f}_{\tau,\sigma}(X) ~=~ \vec{f}_{\tau,\sigma}\qty( \tilde{\sigma}^{-1} \tilde{X} ) ~=~ \vec{f}_{\tau,\, \tilde{\sigma}^{-1}\sigma}\qty( \tilde{X} ).
\end{equation}

Let us note that the UT master integrals by definition in \cref{Tdef} are eigenvectors of the parity conjugation.
Hence, if we evaluated the UT master integrals at a point $X = ( v_1,\ldots, v_5;\,\ep_5)$,
we can obtain their values at the parity-conjugated point $X_{\mathrm{P}} = ( v_1,\ldots, v_5;\,- \ep_5) $
by inverting the signs of the parity-odd integrals,
\begin{subequations} \label{ep5flip}
  \begin{align}
  & \vec{f}^{(+)}_{\tau,\sigma}(X_{\mathrm{P}}) = \vec{f}^{(+)}_{\tau,\sigma}(X),\\
  & \vec{f}^{(-)}_{\tau,\sigma}(X_{\mathrm{P}}) = -\vec{f}^{(-)}_{\tau,\sigma}(X) \,.
  \end{align}
\end{subequations}
The $\mathbb{Z}_2$-grading discussed in \cref{sec_parity,sec:classif} guaranties that also the pentagon functions and the transcendental constants are the eigenvectors of parity conjugation.
Consequently, one can parity-conjugate each function and constant individually in the way that is compatible with \cref{ep5flip}.

Finally, we note that we could restrict our attention to even smaller portion of the physical phase space than $P_0$. Indeed, the $s_{12}$-channel is invariant under the ${\cal S}_2 \times {\cal S}_3$-permutations, which preserve signs of the Mandelstam invariants in \p{s12channel}. Then \cref{eq:mi-any-channel} at $\tilde\sigma \in {\cal S}_2 \times {\cal S}_3 $ and ${\rm sgn}(\tilde \sigma) = +1$ relates all UT master integrals evaluated at a pair of points of $P_0$. If ${\rm sgn}(\tilde \sigma)=-1$, then \cref{eq:mi-any-channel} should be supplemented with \cref{ep5flip}. Thus, knowing the values of the master integrals in the region $P_0 / {\cal S}_2 \times {\cal S}_3 $, which is six times smaller than $P_0$, we can reconstruct values of the master in the whole $P_0$, and consequently, in the whole physical phase space.

In conclusion, we reduced the problem of evaluating the master integrals in arbitrary physical channel to evaluating their permutations in the $s_{12}$-channel. 
Our classification of the pentagon functions in the $s_{12}$-channel is thus sufficient to evaluate the master integrals in an arbitrary physical channel.

\section{Behavior near the boundary \texorpdfstring{$\Delta=0$}{Delta=0}}
\label{sec:delta0}
The obtained analytic expressions for the pentagon functions, enable us to evaluate master integrals in the physical channels and also to study their asymptotic regimes. 
In \cite{Caron-Huot:2020vlo}, the multi-Regge asymptotics in the non-planar sector of five-particle massless amplitudes has been studied. 
One could also study soft or collinear asymptotics by approaching $s_{ij} = 0$ boundaries of the physical scattering channels. 
We are not going to plunge here into the detailed study of all possible singular regimes of the master integrals. 
Instead, we consider asymptotic behaviour of the master integrals (pentagon functions) when approaching $\Delta = 0$ boundary of $P_0$. 
The surface $\Delta = 0$ separates any physical channel into to halves: $\delta >0$ and $\delta <0$. 
It was demonstrated in \cite{Caron-Huot:2020vlo} that the five-particle non-planar Feynman integrals have discontinuities on subvarieties of $\Delta = 0$ and can even be divergent there.
This is a peculiar feature of the non-planar five-particle scattering, which does not manifest itself in simpler planar master integrals studied in the past. 
To gain more experience with the non-planar master integrals we consider $\Delta \to 0$ asymptotics of the pentagon functions.

We should stress that discontinuities and divergences at $\Delta = 0$ appear in the Feynman integrals, but the scattering amplitudes are expected to be free of these singularities in the physical region. In other words, only certain combinations of the Feynman integrals are allowed to contribute to the physical amplitude. The superamplitudes presented in \cite{Caron-Huot:2020vlo} have been tested to satisfy this property. One could try to reverse the argumentation and apply the bootstrap approach to amplitudes \cite{Dixon:2011pw,Dixon:2013eka,Dixon:2020cnr,Chicherin:2017dob}. In the spirit of the Steinmann relations \cite{Caron-Huot:2016owq,Dixon:2016nkn} for the planar hexagon scattering in ${\cal N} = 4$ super-Yang-Mills theory, one could exploit the absence of discontinuities at $\Delta = 0$ as a nontrivial dynamical input on an amplitude ansatz consisting of the pentagon functions. It would be interesting to see how strong is this restriction, and to which extent it fixes nonplanar five-point two-loop amplitudes, in particular the QCD helicity amplitudes.

We choose a generic kinematic point $X_b$ on the boundary $\Delta = 0$ of $P_0$ \p{s12channel}. 
It describes a configuration of momenta $\{ p_i^\mu \}_{i = 1}^5$ lying in a 3-dimensional hyperplane, 
but none of the momenta are soft or collinear, i.e.\ none of the Mandelstam invariants $s_{ij}$ vanish,
\begin{align}
X_b: \qquad s_{12},s_{34},s_{35},s_{45} > 0 
\;,\quad s_{13},s_{14},s_{15},s_{23},s_{24},s_{25} < 0 \;, \quad
\Delta = 0 \,. \label{Delta=0reg}
\end{align}   
One can easily check that all parity-odd letters are equal to 1 at $X= X_b$,
\begin{align}
  a_{k}(X_b) \neq 0 \;, \quad W_{25+ k}(X_b) = 1 \qq{for} k = 1,\ldots,5\,, \label{Wodd1}
\end{align}
and the statement does not depend on the path%
\footnote{The statement does not apply to non-generic points on the $\Delta = 0$ surface for which one or several $s_{ij}$ vanish simultaneously.}
we choose to approach $X_b$.

Let us inspect the pentagon functions in the asymptotic regime $X\to X_b$.
The results of this subsection are implemented in the Mathematica package (see \cref{sec:math_package}).

\subsection{Weights one and two}

We start with weights one and two. In view of \cref{Wodd1}, the parity-odd functions in \cref{g21,itintodd1,itintodd2} take the following form
\begin{subequations} \label{w12onDelta}
  \begin{align}
  & g^{(2)}_{2,i}(X_b) = 0 \,, && \qq{for} i = 1,\ldots,9 \,, \\
  & g^{(1)}_{3,k}(X_b) = - 2 \ii\pi\,\theta(-a_{k}(X_b) ) - \ii\pi \,\delta_{0 a_{k}(X_b)} + \frac{\ii\pi}{3} \;, && \qq{for}  k = 1,2,3,  \\
  & g^{(1)}_{3,k}(X_b) = 2 \ii\pi\,\theta(a_{k}(X_b)) + \ii\pi \,\delta_{0 a_{k}(X_b)} - \frac{2\ii\pi}{3} \;, && \qq{for}  k = 4,5 \,, 
  \end{align}
\end{subequations}
while among the parity-even pentagon functions in \cref{g11,g12,g21} only the form of $g^{(1)}_{2,10}(X) = \log W_{31}(X)$ changes at $X \to X_b$: it is divergent on $\Delta = 0$. 

\begin{figure}[ht]
  \centering
  \includegraphics[width=0.4\textwidth]{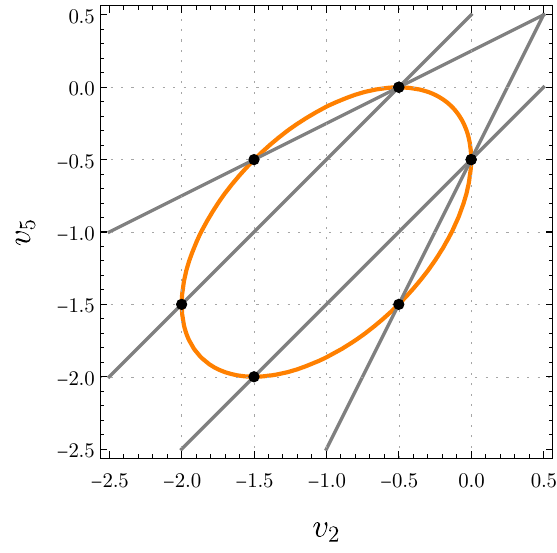}
  \caption{
    Section of the physical region $P_0$ by the plane  $v_1 = 3$, $v_3 =v_4 = 1$.
    The physical region is inside the orange ellipse depicting $\Delta = 0$. 
    The four gray lines, representing $a_{k} = 0$ with $k=1,3,4,5$ (see \cref{a_def}),
    split the surface $\Delta = 0$ in six arcs. 
    The functions $g^{(1)}_{3,k}$, $k=1,3,4,5$, are constant on each arc.
  }
  \label{fig:secphysregion}
\end{figure}

Let us note that $g^{(1)}_{2,10}$ is absent in the weight-1 \p{solf1_g11} and weight-2 \p{solf2_g_even}, \p{solf2_g_odd} solutions of the DE. Thus, they are finite at $\Delta = 0$. 
Approaching the surface $\Delta = 0$ from the opposite sides --- $\delta >0$ and $\delta <0$ --- inside any physical channel,
we find a discontinuity in the parity-odd UT master integrals if they do not vanish at $\Delta = 0$. 
Inspecting the parity-odd weight-2 solutions given in \cref{solf2_g_odd} at $\Delta = 0$,
which involve only weight-1 pentagon functions $g^{(1)}_{3,k}(X_b)$, we find that they are not identically zero \cite{Caron-Huot:2020vlo}. 
More precisely, they vanish on some parts of $\Delta = 0$ carved out by $\{ a_{k}(X)= 0 \}_{k=1,3,4,5}$, while are constant and nonzero on the remaining parts.
This is illustrated in \cref{fig:secphysregion}.

\subsection{Weight three}

Let us find asymptotics of the weight-3 pentagon functions at $\Delta \to 0$ using their integral representation in \cref{w3onefold}. 
The integration path $\gamma$ is a line segment \p{Xt} with the end point $X(t=1) = X_b$. 
The functions $h^{(2)}(X(t))$ from \cref{w3onefold} are finite at $t = 1$. 
On the line segment $\gamma$, we find (see \cref{Wodd1,pent_alph}) that
\begin{equation}
  \ep_5(t) =  {\cal O}(\sqrt{1-t}), \qquad \log W_{k}(t) = {\cal O}(\sqrt{1-t}), \qfor k = 26,\ldots,30.
\end{equation}
Then the integration kernels in \cref{w3onefold} involve the following singularities at $t \to 1$:
\begin{subequations}
  \begin{align}
    \dd \log W_{31}(t) &= {\cal O}\left( \frac{1}{1-t}\right) \dd{t}\,, \\
    \dd \log W_{k}(t) &= {\cal O}\left( \frac{1}{\sqrt{1-t}}\right) \dd t, \qfor k = 26 ,\ldots, 30 .
  \end{align}
\end{subequations} 
The latter are integrable singularities, and only $\dd \log W_{31}(t)$ introduces a divergence.
We regularize it with $\varepsilon \to 0$ as follows,
\begin{multline} \label{regw3}
{\rm reg}\qty[\int\limits^1_0 \dd{t} \, \pa_t \log(W_{31}(t))\cdot f(t)] = \\
  \int\limits^{1}_0 \dd{t} \, \pa_t \log\qty(W_{31}(t))\cdot (f(t)-f(1)) ~+~ f(1) \int\limits^{1-\varepsilon}_0 \dd{t} \, \pa_t \log\qty(W_{31}(t))\,,
\end{multline}
where $f(t)$ is a regular function, and the last term contains logarithmic divergence if $f(1) \neq 1$,
\begin{align}
\int\limits^{1-\varepsilon}_0 \dd{t} \, \pa_t \log{(W_{31}(t))} = \log{(\ep_{5}(1-\varepsilon))}-\log{(\ii\sqrt{3})}\,.
\end{align}
Inspecting all weight-3 pentagon functions $\{ g^{(3)}_{i} \}_{i=1}^{111}$ \p{w3onefold}, we find that only parity-odd $g^{(3)}_{98}$ and $g^{(3)}_{103}$ are singular at $\Delta = 0$.

Parity-odd weight-3 solutions \p{solw3_odd} of the DEs involve logarithmically divergent at $\Delta \to 0$ weight-1 and weight-3  pentagon functions $g^{(1)}_{2,10}$, $g^{(3)}_{98}$, $g^{(3)}_{103}$,
and indeed we observe divergences of the UT master integrals at $\Delta = 0$.

\subsection{Weight four}

In a similar spirit, we study asymptotics of the weight-4 pentagon functions given in \cref{w4onefold2}. 
Only terms with $k=31$ and/or $j = 31$ produce singularities at $X = X_b$. 
Thus, we need to regularize three types of terms:
\begin{subequations}
  \begin{align}
  & \int^1_0 \dd{u} \, \pa_u \log(W_{31}(u))\cdot f(u) \,,\label{term1}\\
  & \int^1_0 \dd{u} \int^1_u \dd t\, \pa_t \log(W_{31}(t)) \cdot f(u) \,,\label{term2}
  \\
  & \int^1_0 \dd{u} \, \pa_u \log(W_{31}(u))  \int^1_u \dd t \,\pa_t \log(W_{31}(t)) \cdot f(u), \label{term3}
  \end{align}
\end{subequations}
where $f(u)$ is a regular function. 
We have already regularized the first term \p{term1} in \p{regw3}. 
Taking into account \cref{intdlogw31}, we regularize the second term, \cref{term2}, as follows:
\begin{multline} \label{regterm2}
 {\rm reg} \qty[\ref{term2}]  =  \int\limits^{1-\varepsilon}_0 \dd{u} \int\limits^{1-\varepsilon}_u \dd t\, \pa_t \log(W_{31}(t)) \cdot f(u) = \\
 =  \log(\ep_5(1-\varepsilon)) \int^1_0 \dd{u}\, f(u) - \int^1_0 \dd{u}\, \log(\ep_5(u)) \,f(u)\,.
\end{multline}
Let us note that $\log{(\ep_5(u))}$ is an integrable singularity at $u \to 1$. 
Finally, we regularize the third term \p{term3} as follows:
\begin{multline} 
\mathrm{reg} \qty[\ref{term3}] =  \int\limits^{1-\varepsilon}_0 \dd{u} \, \pa_u \log(W_{31}(u)) \,  \int\limits^{1-\varepsilon}_u \dd t \,\pa_t \log(W_{31}(t)) \cdot f(u) =  \\
  - \int^1_0 \dd{u}\,  \log (\ep_5(u)) \,\pa_t \log(W_{31}(t))\cdot (f(u)-f(1)) \\
 + \log{(\ep_5(1-\varepsilon))} \int^1_0 \dd{u}\,   \pa_u \log{(\ep_5(u))} \cdot(f(u)-f(1)) \\
 + \frac{f(1)}{2}\left(\log\qty(\ep_5(1-\varepsilon))-\log{(\ii\sqrt{3})} \right)^2 \label{regterm3}
\end{multline}
The integrations in \cref{regterm3} are convergent, and singularities are revealed in the form of divergent $\log{(\ep_5(1-\varepsilon))}$ as $\varepsilon \to 0$. 

The divergent terms in \cref{term1,term2,term3} are present only in the parity-odd pentagon functions of weight four. 
Thus, only they contain logarithmic divergences $\log \ep_5$ and $\log^2 \ep_5$ in the limit $\Delta \to 0$. 
These divergences of the pentagon functions are also inherited by the parity-odd weight-4 solutions of the DEs given by \cref{solw4_odd}.

\section{Numerical Evaluation}
\label{sec:numerical}

In \cref{sec:classif},
we constructed a complete set of pentagon functions required to analytically represent (up to weight 4) all UT master integrals of the topologies shown in \cref{fig:integral_families}.
In this section, we describe our implementation of numerical evaluation of the pentagon functions and the UT master integrals.

All results of \cref{secAnyChannel,sec:delta0} are implemented in a \texttt{Mathematica} package, discussed in \cref{sec:math_package}.
The package is provided mainly for the purpose of demonstration,
and it is not intended for the use cases where high throughput and/or numerical robustness is required.
For the later, we provide a \texttt{C++} library, which we present in \cref{sec:cpplib}.
The library is optimized for performance,
and its numerical efficiency makes it well-suited for evaluation of phase-space integrals with the Monte-Carlo method ---
the key ingredient for obtaining theoretical predictions for any observable cross section.

\subsection{\texttt{Mathematica} package \texttt{PentagonMI}}
\label{sec:math_package}
The \texttt{Mathematica} package \texttt{PentagonMI} implements numerical evaluation
of all UT master integrals $\vec{f}_{\tau,\sigma}$, defined in \cref{changebasis,UTint_eps_exp,UTint1loop_eps_exp},
at any point of the physical phase space. The master integrals are expressed in the basis of the pentagon functions, 
constructed in \cref{sec:classif}. The package consists of three main components.
The first component is data files containing the definitions of the objects employed in this paper.
The data files are in \texttt{Mathematica} format.
However, they are made to be self-consistent and understandable as plain text, such that they can be used outside of this package.\cprotect\footnote{%
  The file \verb|datafiles/constants_numerical.m| is an exception.
}
The second component uses the definitions of master integrals from \verb|datafiles/| to construct their analytic expressions
in terms of pentagon functions. The third component implements numerical evaluation of the pentagon functions.
The package can be obtained from the \texttt{git} repository \cite{PentagonMI} with
\begin{lstlisting}
  git clone https://gitlab.com/pentagon-functions/PentagonMI.git
\end{lstlisting}
To install the package, follow the instructions in the ``Installation'' section of the \texttt{README.md} file in the root directory of 
the distribution.

\begin{table}[t]
  \centering
  \begin{tabular}{Ll}
    \toprule
    \mathbf{P} &  pentagon  \\
    \mathbf{PB} &  planar pentagon-box  \\
    \mathbf{HB} &  hexagon-box  \\ 
    \mathbf{DP} & double pentagon \\
    \bottomrule
  \end{tabular}
  \caption{The abbreviations for the Feynman integral topologies shown in \cref{fig:integral_families}, which are used by the interface of \texttt{PentagonMI}.}
  \label{tab:topoabbr}
\end{table}

Let us first describe the data files in the directory \verb|datafiles/| provided with the package.
Our choice of UT master integrals, i.e.\ \cref{changebasis},
is specified for each Feynman integral topology by a corresponding file in the directory \verb|datafiles/UT-MI/| (see abbreviations of topologies $\tau$ in \cref{tab:topoabbr}).
To reduce the size of files,
we write all UT master integrals as $\mathbb{Q}$-linear combinations of a smaller subset of 1917 UT integrals $\{ \mathbf{G} \} \subset \{ \vec{f}_{\tau,\sigma} \}$ as
\begin{equation} \label{eq:mi-in-g}
  \vec{f}_{\tau,\sigma}(X) = \sum_{i} \qty(\vec{c}_{\tau,\sigma})_i ~\mathbf{G}_i(X),
\end{equation}
which we found by identifying Feynman integrals among different topologies and permutations \cite{Badger:2019djh}.
Each UT master integral is rewritten as a linear combination of UT integrals $\mathbf{G}$ in the file \verb|datafiles/MI_in_G.m|.
UT integrals $\mathbf{G}$ expressed in terms of pentagon functions, as given by \cref{solf1_g11,solf2_g_even,solf2_g_odd,solw3_even,solw3_odd,solw4_even,solw4_odd},
can be found in files \verb|datafiles/GtoF_weight*.m|.
The algebraically-independent transcendental constants, shown in \cref{tab_trans_const}, are defined in \verb|datafiles/constants.m|. We provide weight-0 initial values (see \cref{f0sol}) for all four topologies in \verb|datafiles/initial_values_weight0.m|.
These values are invariant under permutations, so we provide them for each topology in a single permutation $\sigma$.
Finally, the definitions of pentagon functions as given in \cref{g11,g12,g13,g21,g22,eq:g33p,eq:g33m,w4onefold2} can be found in files
\begin{lstlisting}
    datafiles/functions_weight1.m
    datafiles/functions_weight2.m
    datafiles/functions_weight3_onefold.m
    datafiles/functions_weight4_onefold.m
\end{lstlisting}
The parity grading of the alphabet, master integrals, pentagon functions, and transcendental constants plays an important role in our classification.
We list all parity-odd objects in the file \verb|datafiles/parity-grading.m|.
Further details can be found in the \verb|datafiles/README.md| file supplied along with the distribution.

The main interface of \texttt{PentagonMI} is given by the function \texttt{EvaluateMI}, which accepts a list
of master integrals to be evaluated, and a kinematical point in the physical region given by the five Mandelstam invariants in \cref{eq:kinematics}.
The master integrals are indexed according to their definitions found in the directory \verb|datafiles/UT-MI/|.
A UT master integral $\qty(\vec{f}_{\tau,\sigma})_i$ is identified in \texttt{PentagonMI} by its topology abbreviation (see \cref{tab:topoabbr}),
the index of permutation $\sigma$ taking integer values from $1$ to $120$ and defined in the file \verb|datafiles/permutations.m/|, and the index $i$ of the UT integral within the given family.
For example,
\begin{lstlisting}[language=Mathematica]
EvaluateMI[
  {DP[3,108], HB[100,21], PB[120,61]},
  {4, -(113/47), 281/149, 349/257, -(863/541)}
]
\end{lstlisting}
will evaluate the double pentagon integral \#108 in permutation $\sigma = \{ 1,2,4,3,5\}$, which is indexed by 3, 
the hexagon-box integral \#21 in permutation $\sigma = \{ 5,1,3,4,2\}$, which is indexed by 100, and the pentagon-box integral \#61 in permutation $\sigma = \{ 5,1,3,4,2\}$, which is indexed by 120,
at the kinematical point $\{4, -(113/47), 281/149, 349/257, -(863/541)\}$.
The function returns coefficients of the $\epsilon$-expansion (see \cref{UTint_eps_exp,UTint1loop_eps_exp}) of each UT master integral.
If the kinematical point does not belong to the $s_{12}$-channel $P_0$ \eqref{s12channel}, 
\texttt{EvaluateMI} uses \cref{eq:mi-any-channel} to find a permutation that
maps it to $P_0$. 
By default, $\delta>0$ is assumed, and evaluation of master integrals at a parity-conjugated point ($\delta < 0$) can be requested with the option
\verb|"ParityConjugation" -> True|.
Several other options that can be used to modify certain aspects of evaluation are available;
we refer to the documentation provided in the file \verb|PentagonMI.m|.
For an example, see the program \verb|test/all_master_integrals.m|, which evaluates all master integrals in all permutations at a single phase-space point. 

\texttt{EvaluateMI} is only responsible for constructing a representation of each UT master integral in terms of pentagon functions.
In order to obtain numerical values of the master integrals, numerical values of the pentagon functions at the given
kinematical point are required.
Numerical evaluation of the pentagon functions is carried out either with the (sub-)package \texttt{PentagonFunctionsM}, described in the next section,
or through a \texttt{Mathematica} interface of the \texttt{C++} library (see \cref{sec:cpplib}).
By default, the latter is chosen if available, and the option \verb|"UseCppLib" -> False| can be used to choose
the \texttt{Mathematica} implementation instead.

\subsubsection*{Numerical evaluation of pentagon functions}

We implemented numerical evaluation of pentagon functions in a \texttt{Mathematica} package \texttt{PentagonFunctionsM}.
The package can be used independently or as a part of \texttt{PentagonMI}.

All functions are evaluated in the analyticity region $P_0$ \eqref{s12channel}.
We evaluate weight-1 and weight-2 functions, explicitly given by \cref{g11,g12,g13,g21,g22}, using
the standard \texttt{Mathematica} functions \texttt{Log} and \texttt{PolyLog}.
For weight-3 and weight-4 functions, we use one-fold integral representations in \cref{eq:g33p,eq:g33m,w4onefold2}.
We carry out the numerical integration with the built-in \texttt{Mathematica} function \texttt{NIntegrate}.

The pentagon functions are represented as \verb|F[w,i,j]| at weights $w=1,2$ and
as \verb|F[w,i]| at weights $w=3,4$, where indices $i,j$ are in one-to-one correspondence with the indices of \cref{g11,g12,g13,g21,g22} and
\cref{eq:g33p,eq:g33m,w4onefold2} respectively.
Numerical values of a list of functions can be obtained by calling the function \texttt{EvaluatePentagonFunctions}.
For example,
\begin{lstlisting}[language=Mathematica]
EvaluatePentagonFunctions[
  {F[1,3,1], F[3,17], F[4,113], F[4,470]},
  {4, -(113/47), 281/149, 349/257, -(863/541)}
]
\end{lstlisting}
evaluates the pentagon functions $g^{(1)}_{3,1},\,g^{(3)}_{17},\,g^{(4)}_{113},\,g^{(4)}_{470}$.
The requested numerical integration error of weight-3 and weight-4 functions can be set with the option \verb|"IntegrationPrecisionGoal"|.
Its default value is 10, which means that the integration is terminated when the (negative) $\log_{10}$ of an estimate of either relative or absolute error
reaches 10. The requested functions are evaluated in parallel by default, using all available CPUs. Parallelization can be disabled by setting
the option \verb|"Parallel" -> False|.

Kinematical points are allowed to lie on the surfaces of spurious singularities exactly.
In this case, the corresponding $\dd{\log}$-kernels are set to zero, see discussion around \cref{badWt,eq:g33m,eq:intdlogsp}.

As a special case, \texttt{EvaluatePentagonFunctions} also  evaluates asymptotics of the pentagon functions at $\Delta\to 0$ as
discussed in \cref{sec:delta0}. The special case is activated automatically whenever evaluation at a kinematical point sitting on the boundary $\Delta=0$ is requested.
Concretely, for weight-3 and weight-4 pentagon functions, we implemented the regularization of the divergent one-fold integrals as introduced in \cref{regw3,regterm2,regterm3}.
The asymptotics, which is divergent in the limit $\Delta\to 0$, is (at most quadratic) polynomial in $\log(\ep_5)$ with numerical coefficients resulting from the regularized one-fold integrations.
An example can be found in the test program \verb|test/functions_delta_singular.m|.
In this program all pentagon functions are evaluated at a kinematical point $X_b$ with $\Delta(X_b)=0$.
Also, as a consistency check, asymptotics of the divergent at the point $X_b$ weight-4 functions are compared to their
values at a point that is slightly deformed away from $\Delta=0$.

\subsection{\texttt{C++} library \texttt{PentagonFunctions++}}
\label{sec:cpplib}
One of the main goals of this paper is to take advantage of analytic understanding
of the five particles massless scattering to derive a representation of
the corresponding two-loop master integrals that is suitable for phenomenological applications.
In particular, this representation should lend itself to a numerically efficient and stable implementation.
We believe that the classification of pentagon functions, which we carried out in \cref{sec:classif}, indeed provides such a representation.
Nonetheless, we find that the \texttt{Mathematica} implementation described in the previous section does not realize the full potential of our method.
To this end, we implement numerical evaluation of pentagon functions in a \texttt{C++} library \texttt{PentagonFunctions++},
which we present in this section.

\subsubsection{Features}

\texttt{PentagonFunctions++} is a \texttt{C++14} library, which implements numerical evaluation of the pentagon functions, classified in \cref{sec:classif},
in their analyticity region $P_0$ \eqref{s12channel}. 

For numerical evaluation of weight-1 and weight-2 functions, we use their explicit representation in \cref{g11,g12,g13,g21,g22} in terms of the $\log$, $\Li_2$, and $\mathrm{Cl}_2$ functions.
The latter are evaluated numerically with a custom \texttt{C++} implementation \cite{Li2pp} based on the algorithms of \cite{vanHameren:2005ed}. 
For weight-3 and weight-4 functions, we use the one-fold integral representations in \cref{eq:g33p,eq:g33m,w4onefold2} and evaluate the integrals numerically.
The integrands of certain weight-3 and weight-4 functions are somewhat lengthy.
Thus, to speed up their numerical evaluation, we optimize the integrands with respect to the number of floating-point operations with
the code-optimization facilities \cite{Kuipers:2013pba} of the computer algebra system \texttt{FORM} \cite{Ruijl:2017dtg}.\footnote{%
  We remark that this optimization can potentially be in conflict with numerical stability, see the discussion in \cref{sec:cpp-perf}.
}

The choice of a numerical integration algorithm for the evaluation of the one-fold integrals (\emph{quadrature}) can significantly impact evaluation times.
Thus, it is essential to choose an algorithm that is suitable for the problem at hand.
We employ the double exponential \emph{tanh-sinh} quadrature \cite{Mori:1973}. 
The quadrature exploits a change of an integration variable $t \in (0,1)$,
\begin{equation}
  t = \frac{1}{2} \qty(1 + \tanh(\frac{\pi}{2} \sinh(x))),
\end{equation}
which maps the endpoints of the integration region to infinities, $x\in (-\infty,+\infty)$,
and the transformed integrand decays double exponentially, i.e.\  
as $\exp\qty(-\frac{\pi}{2}\exp(\abs{x}))$ with $x\to \pm \infty$. It can then be shown \cite{Mori:1973} that
the integral can be approximated remarkably well by a simple trapezoidal rule.
In fact, it was proven in \cite{Mori:1973, Bailey:2005} that the tanh-sinh quadrature
is the optimal choice for integrands that are analytic inside the integration domain (excluding, perhaps, the endpoints)
in a sense that it requires the least number of evaluations of the integrand to reach a given integration error.
For this class of integrands, the tanh-sinh quadrature converges exponentially, i.e.\ the number of correct digits 
in the numerical approximation is proportional to the number of evaluations of the integrand.
Integrable singularities at the endpoints of the integration domain,
such as the logarithmic singularity in the integrands of the weight-4 functions (see \cref{sec:w4onefold}),
do not introduce any complications for this quadrature, hence no special handling is required.
\texttt{PentagonFunctions++} uses an adapted implementation of the tanh-sinh quadrature from \texttt{Boost C++} \cite{BoostTanhSinh}.

We pointed out in \cref{badletters,badWt} that several letters $W_{k\in \alpha}$ of the pentagon alphabet do not have a definite sign inside the analyticity region $P_0$
and vanish at the endpoint $t=0$ of the integration interval. Thus, their $\dd{\log}$-forms have a simple pole at $t=0$.
As we discussed around \cref{eq:vanishing-dlog-w3,eq:vanishing-dlog-w4}, these poles are compensated by vanishing combinations of weight-2 functions.
The quadrature algorithm discussed above might require evaluation of the integrands very close to the endpoints.
It is thus important to ensure that the cancellation of the poles is numerically stable.
To this end, in the neighborhood of $t=0$, $t <\tilde{t}$  we evaluate the kernels $\dd{\log\qty(W_k(t))}$ together with their coefficients $h_k(t)$ through
their generalized series expansion around $t=0$ as
\begin{equation}
  h_k(t) \derivative{\log(W_k(t))}{t} \xrightarrow{t < \tilde{t} \ll 1} h^{0,0}_k + h^{0,1}_{k} \;\log(t) + h^{1,0}_k\;t + h^{1,1}_k \; t \log(t) + \mathcal{O}(t^2),
\end{equation}
such that no numerical cancellation of the pole has to occur.
The threshold  $\tilde{t}$ is chosen such that 
$\tilde{t}^2 \ll \varepsilon_\text{T}$, where $\varepsilon_\text{T}$ is the roundoff error (or \emph{machine epsilon}) of the
floating-point number type $\text{T}$ (e.g.\ $\varepsilon_{\text{double}} \simeq 10^{-16}$).

On certain subvarieties of the physical phase space (\emph{spurious singularities}) the letters $W_{k \in \alpha}$ might be identically zero.
As we mentioned around \cref{badWt}, on these subvarieties the corresponding integration kernels also vanish.
However, in small neighborhoods of these subvarieties the letters $W_{k \in \alpha}(\gamma(t)) = b_k(X)\,t$ \emph{almost} vanish along the whole line segment $\gamma=[X_0;X]$,
i.e.\ $0 < \abs{b_k(X)} \ll 1$.
Then the contribution from $h_k \dd{\log W_{k}}$ to the integral is rendered small by potentially large cancellations in $h_k(t)$.
In principle, this situation can be avoided by using an appropriate representation of $h_k(t)$ and/or path deformations.
However, we find that it is sufficient to simply set the integration kernels $\dd{\log\qty(W_k)}$ to zero exactly whenever $\abs{b_k(X)}$ is below
a certain threshold. We note that the pentagon functions are analytic on the surfaces of spurious singularities,
and only the functions that vanish on a particular spurious-singularity surface can be significantly impacted by this procedure. 
The threshold can thus be adjusted in such a way that only insignificant neighborhoods of the spurious-singularity surfaces are potentially affected. 
We demonstrate  this \textit{a posteriori} in \cref{sec:cpp-perf}.
We leave a more refined analysis of the spurious singularities for future study.

\texttt{PentagonFunctions++} is able to perform all evaluations in three fixed-precision floating-point types: 
double, quadruple and octuple precision, which  respectively represent significands of approximately 16, 32, and 64 decimal digits.
We use a \texttt{C++} implementation of quadruple and octuple numerical types from the \texttt{qd} library \cite{QDlib}.
Numerical evaluation in multiple fixed-precision types is indispensable for understanding numerical stability of the implementation,
as well as for adaptively balancing precision against performance.

\subsubsection{Usage}

The library can be obtained from the \texttt{git} repository \cite{PentagonFunctions:cpp} with
\begin{lstlisting}
  git clone https://gitlab.com/pentagon-functions/PentagonFunctions-cpp.git 
\end{lstlisting}
To install the library, follow the instructions in the ``Installation'' section of the \texttt{README.md} file in the root directory of the distribution \cite{PentagonFunctions:cpp}.

The intended way to use \texttt{PentagonFunctions++} is to write a \texttt{C++} program, which links to the provided static or shared library.
Further details can be found in the ``Usage'' section of the \verb|README.md| file found in the root directory of the distribution \cite{PentagonFunctions:cpp}.

The main interface of the library is provided by a \verb|struct| \verb|FunctionID|, which is declared in the header file \verb|src/FunctionID.h|.
An instance of \verb|FunctionID|, constructed with integer arguments \verb|(w,i,j)| or \verb|(w,i)|, represents the pentagon function of weight \verb|w|  and indices
\verb|i,j|, according to the definitions in \cref{g11,g12,g13,g21,g22,eq:g33p,eq:g33m,w4onefold2}.
The instances of \texttt{FunctionID} can be used to obtain callable \emph{function objects} of numerical type \texttt{T} (e.g.\ \verb|double|, \verb|dd_real|, \verb|qd_real|) with
the method \verb|get_evaluator<T>()|. For example, with
\begin{lstlisting}[language=C++]
  FunctionID fid{4,471};
  auto f = fid.get_evaluator<double>();
\end{lstlisting}
one creates a function object \verb|f|, which can be used to numerically evaluate the pentagon function $g^{(4)}_{471}$ at any number of kinematical points in double precision.
The method \verb|get_evaluator<T>| performs an initialization stage of the integration framework that need not be repeated for each subsequent evaluation.
Several example programs can be found in \verb|examples/| directory of the distribution.

The termination condition of the numerical integration is controlled by the global variable 
\begin{lstlisting}[language=C++]
template <typename T> extern T IntegrationTolerance;
\end{lstlisting}
which specifies the tolerance for each numerical type \verb|T| independently.
It is declared in the header file \verb|src/Constants.h|.
The numerical integration is terminated when the difference of two subsequent estimates of the integral have absolute value less than the tolerance multiplied
by an estimate of the $L^1$ norm of the integral.
The default value is chosen such that the integration error is close to the rounding error of the numerical type \verb|T|.
Finer control over tolerance might be exploited for improving either integration speed or precision of the results.

For convenience, we also provide a \texttt{Mathematica} interface.
It is realized as a \texttt{Mathematica} package \texttt{PentagonFunctions},
which interacts with the program\\ \verb|mathematica_interface/evaluator_math.cpp|.
The interface is similar to the one of the package \texttt{PentagonFunctionsM}, described in the previous section.
An example of the interface usage can be found in \verb|examples/math_interface.m|.

\subsubsection{Performance}
\label{sec:cpp-perf}

In this section, we demonstrate performance of our implementation with respect to evaluation speed and numerical stability,
which are the most important properties of a numerical algorithm.

To characterize evaluation speed of our implementation,
we evaluate all pentagon functions with \texttt{PentagonFunctions++} (with the standard settings) 
at a random generic point from the physical phase space.  We perform the evaluation on a single core 
of \textit{Intel(R) Core(TM) i7-7700 CPU @ 3.60GHz}. We show the evaluation times as well as the
(minimal) number of correct digits \footnote{%
  The target values are obtained with the package \texttt{PentagonFunctionsM}, see \cref{sec:math_package}.
} for the three supported floating-point types in \cref{tab:timings}.
Evaluation of the planar subset of the pentagon functions\footnote{%
The subset of the pentagon functions contributing to master integrals of the pentagon and the planar pentagon-box topologies.
}
takes approximately $40\%$ of the total evaluation time.
Comparing to the evaluation times of the planar pentagon functions of \cite{Gehrmann:2018yef} reported in \cite{Chawdhry:2019bji},
we observe that the planar subset of our implementation evaluates approximately 100 times faster.

\begin{table}[ht]
  \centering
  \begin{tabular}{ccc}
    \toprule
    Precision & Correct digits & Timing (s) \\
    \midrule
    double & 13  & 2.5 \\
    quadruple & 29 & 180 \\
    octuple & 60 & 3900  \\ 
    \bottomrule
  \end{tabular}
  \caption{Evaluation times of all pentagon functions on a typical phase-space point. Evaluation is performed in a single thread.}
  \label{tab:timings}
\end{table}

\begin{figure}[t]
  \centering
  \includegraphics[width=0.8\textwidth]{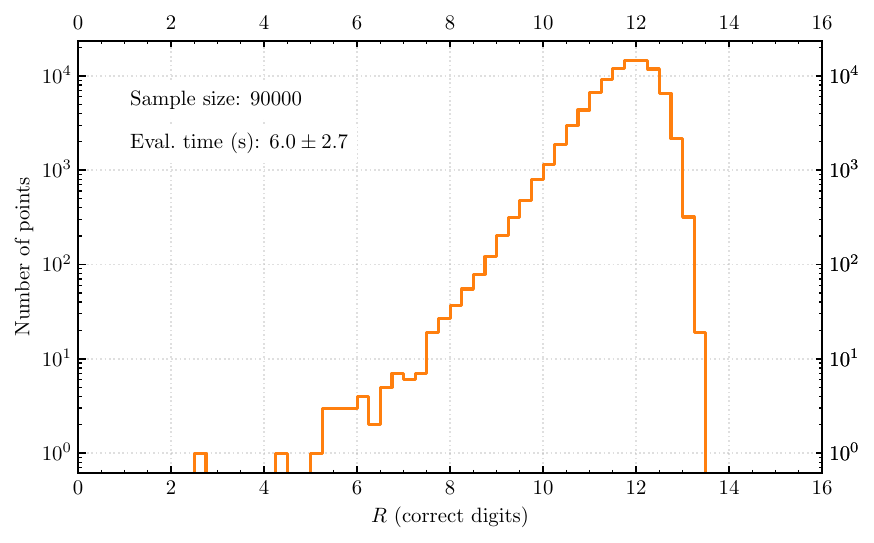}
  \caption{
    Histogram of minimal logarithmic relative error of pentagon functions (see \cref{eq:mindigits}) sampled on 90000 kinematical points of a generic five-particle physical phase space.
    The average evaluation time in double precision is obtained from running 64 parallel jobs on a server with \textit{Intel(R) Xeon(R) Silver 4216 CPU @ 2.10GHz}.
  }
  \label{fig:num_stab}
\end{figure}

Further, we demonstrate the numerical stability of our implementation by evaluating all pentagon functions on a sample of 90000 phase-space points,
drawn from a typical distribution employed in computations of differential cross sections for processes with five massless particles \footnote{%
  More concretely, we use an integration grid, optimized for Monte-Carlo integration of
  the leading order $q\bar{q} \to \gamma\gamma\gamma$ matrix elements over
  the fiducial phase space defined by the analysis of \cite{Aaboud:2017lxm}.
  We used \texttt{MATRIX} \cite{Grazzini:2017mhc} to obtain the integration grid.
}.
We evaluate all pentagon functions in double and quadruple precision at each phase space point, and we use the latter to compute the accuracy of the former.
We characterize the accuracy of the evaluation $\hat{g}_i(X)$ of the $i$-th pentagon function on a kinematical point $X$ by the logarithmic relative error (``correct digits'') $r_i(X)$ which we define as
\begin{equation}\label{eq:digits}
  r_i(X) = -\log_{10} \left| \frac{\hat{g}_i(X) - \hat{g}^{\text{(q)}}_i(X)}{\hat{g}^{\text{(q)}}_i(X)} \right|,
\end{equation}
where $\hat{g}^{\text{(q)}}_i(X)$ is the numerical evaluation of the same function in quadruple precision.
We define the \emph{minimal} logarithmic relative error among all pentagon functions at the kinematical point $X$ as
\begin{equation} \label{eq:mindigits}
  R(X) = \min_i[r_i(X)], \qquad i \in \{\text{all pentagon functions}\} .
\end{equation}
We display the distribution of $R(X)$ over the phase space in \cref{fig:num_stab}.
We observe very good numerical stability in the bulk of the phase space:
only $0.1\%$ of the phase-space points evaluate with less than 8 correct digits.

All 12 kinematical points $\vb{X}_{(R<6)}$ with $R<6$ are from the region of the phase space with $ 0 < \delta \ll 1$ .
As we discussed in \cref{sec:delta0}, some pentagon functions diverge in the limit $\delta\to 0$.
But the divergence is only logarithmic. So, with $\min_{\vb{X}_{(R<6)}}[\delta] \gtrsim 10^{-7}$, the absolute values of the divergent pentagon functions still remain relatively small.
However, the condition number $\kappa$ of e.g.\ the function $g^{(1)}_{2,10} =\log(\ii\delta)$
diverges much faster,
\begin{equation}
 \kappa(g^{(1)}_{2,10}) \xrightarrow{\delta\to 0} \order{\frac{1}{\delta^2\log^2(\delta)}}.
\end{equation}
In other words, numerical evaluation of the function $g^{(1)}_{2,10}$ in the regime $\delta\ll 1$ becomes dominated by the rounding error of the input data (Mandelstam invariants)
much earlier than the function itself becomes large.
Let us mention that it should be possible to circumvent this issue in applications to numerical evaluation of 
complete two-loop amplitudes expressed in terms of the pentagon functions. We leave these considerations for future studies.

We would like to note that the desired precision of pentagon function evaluation heavily depends on the intended application.
For this reason, we did not attempt to reach any given accuracy threshold.
Instead, we have studied the numerical stability of double-precision evaluation in the default running mode of \texttt{PentagonFunctions++}.
If a certain accuracy threshold is to be guaranteed, the following strategy can be attempted.
Only the kinematical points close to either spurious singularities or $\delta=0$ are expected to be problematic,
which can be detected before any integrations take place.
Then the potentially problematic functions can be evaluated in higher precision with the help of multi-precision facilities of \texttt{PentagonFunctions++}.

We conclude that the performance of our numerical implementation of the pentagon functions will almost certainly be sufficient for 
the main anticipated application in phenomenological studies.

\section{Validation}
\label{sec:validation}
The presented analytic expressions for the master integrals as well as their numerical implementations have passed a number of cross checks which we report in this section.

The classification of all pentagon functions (planar and nonplanar) elaborated in this paper does not directly rely on the planar classification from \cite{Gehrmann:2018yef},
which brings forward the cyclic symmetry.
In other words, the analytic expressions for the master integrals of the planar pentagon-box topology provided here are not literally the same as in \cite{Gehrmann:2018yef},
but of course both should be in agreement.
Indeed, using our code, we reproduce numerical values of the planar integrals calculated by the code from \cite{Gehrmann:2018yef}.
Furthermore, we employed an implementation of numerical unitarity framework \cite{Ita:2015tya,Abreu:2017hqn,Abreu:2018zmy} in \textsc{Caravel} \cite{Abreu:2020xvt} 
to successfully reproduce the numerical evaluations of the leading-color two-loop five-gluon helicity amplitudes in the physical region presented in \cite{Badger:2018gip}.

One of the crucial ingredients in the analytic solution of the DEs for the master integrals are initial values, see \cref{bc}. For the double-pentagon family we use the initial values from \cite{Chicherin:2018old},
which have been already validated independently with \texttt{pySecDec} \cite{Borowka:2017idc,Borowka:2018goh} for several permutations of this topology\footnote{Version two of the preprint \cite{Chicherin:2018old} contains a relative error of $\approx 1 \%$ in the initial values, which was corrected in version three.}. Multi-digit numerical values of the full set of initial values presented in this paper have been already employed in \cite{Caron-Huot:2020vlo, Chicherin:2018yne},
where expected physical properties have been observed. In particular, the weight-3 and weight-4 initial values have been probed in  \cite{Chicherin:2018yne}, where the weight-drop of the five-gluon two-loop hard function has been explicitly verified. This provides an indirect consistency check.

We check our analytic solutions of the DEs comparing their numerical evaluations with the numerical integration of the DEs via generalized series expansions \cite{Francesco:2019yqt}, implemented in \texttt{DiffExp} \cite{Hidding:2020ytt}.
We employed \texttt{DiffExp} to transport the initial values at the reference point \cref{B} to an arbitrary point in the physical region
that can be reached without leaving the analyticity region and compared the result with our evaluations. We found perfect agreement within numerical precision. We should mention that we integrate each permutation of the DE separately and never cross the physical region boundaries. It would be interesting to analytically continue nonplanar topologies from one scattering channel to another with \texttt{DiffExp}.

Also we numerically evaluated a number of the nonplanar master integrals with \texttt{pySecDec} \cite{Borowka:2017idc,Borowka:2018goh} at arbitrary points of the physical phase space and compared the result with our evaluations of the pentagon functions finding a perfect agreement. The latter check is sensitive to the initial values and the DE integration.

In order to verify the one-fold representations of the pentagon functions, see \p{w3onefold} and \p{w4onefold2}, and their numerical implementation, we constructed alternative representations of the weight-three and four pentagon functions. The weight-three functions are written explicitly as polynomials in the polylogarithmic functions, and weight-four functions are one-fold $\dd{\log}$ integrals with the weight-three integrands (see \cref{sec:alternative_representation}). We find that both implementations of the pentagon functions agree numerically. Obviously, the analytic expressions for the pentagon functions --- the one implemented in the public library and the alternative one --- are completely different. They correspond to different rewritings of the iterated integral form of the pentagon functions into a more tangible form. The numerical agreement of the two representations is a strong check for both of them as well as for their numerical implementations.

Last but not least, the two public numerical implementations of the pentagon functions --- the \texttt{Mathematica} package and the \texttt{C++} library --- enables us to control the quadrature accuracy.

\section{Conclusions}
\label{sec:conclusions}

In this paper, we constructed a complete set of transcendental functions, \emph{pentagon functions}, which are sufficient to represent all planar and non-planar master integrals for 
massless five-point two-loop and one-loop scattering amplitudes up to transcendental weight four. Our analytic results are valid over the whole physical phase space.
This was achieved by expressing all external momenta permutations of master integrals in the same set of pentagon functions.
We expressed the pentagon functions of weights one and two in terms of (di-)logarithms, and we constructed one-fold integral representations for the pentagon functions of weights three and four.
We presented an implementation of numerical evaluation of pentagon functions in a public \texttt{C++} library.
The latter was designed to satisfy demands of phenomenological applications.
Thus, for the first time, all massless five-point two-loop Feynman integrals are available for immediate application in computations of fully differential cross sections. 
Together with the ongoing advances in reduction of five-point two-loop amplitudes,
our results pave the way for computation of NNLO predictions for a number of key scattering processes at hadron colliders.
The latter include production of three hard jets, two-photon and three-photon production in association with jets.

The pentagon functions are crucial for finding compact analytic representations of scattering amplitudes as well as for studying their asymptotic limits. 
In fact, the analytic form of amplitudes can be directly reconstructed from their numerical evaluations over finite fields \cite{Peraro:2016wsq}.
This approach has lead to remarkable progress in calculation of planar five-point amplitudes 
\cite{Badger:2018gip,Badger:2017jhb,Abreu:2017hqn,Abreu:2018zmy,Badger:2018enw,Abreu:2018jgq,Abreu:2019odu,Chawdhry:2019bji,Chicherin:2018yne,Abreu:2018aqd,Chicherin:2019xeg,Abreu:2019rpt}.
We expect that our results will open a possibility of extending these methods to non-planar sector.

Our strategy of constructing bases of transcendental functions can be generalized to scattering processes with even larger number of scales, such as five-point processes with massive particles.
It might be also possible to apply our approach for the cases where Feynman integrals evaluate to iterated integrals over more complicated differential forms,
such as modular forms (see e.g.\ \cite{Broedel:2017kkb,Broedel:2017siw,Adams:2017ejb}).
It would be interesting to explore this in the future.
Finally, an interesting question is to compare efficiency of the numerical evaluation of our analytic results to the approach of solving DEs numerically via generalized series expansions \cite{Francesco:2019yqt,Hidding:2020ytt}.

\acknowledgments

We thank Simon Badger, Gudrun Heinrich, Johannes Henn, Thomas Gehrmann, Simone Zoia for insightful discussions. 
We are grateful to Johannes Henn for comments on the manuscript.
This project is supported by the European Research Council (ERC) under the European Union's Horizon 2020 research and innovation programme,
\textit{Novel structures in scattering amplitudes} (grant agreement No.\ 725110).

\appendix

\section{Pentagon alphabet}
\label{App_alph}

We recall the definition of the nonplanar pentagon alphabet \cite{Gehrmann:2015bfy,Chicherin:2017dob}.
We use shorthand notations 
\begin{align}
  a_i \coloneqq v_i v_{i+1} + v_{i+2} v_{i+3} - v_{i+1} v_{i+2} - v_i v_{i+4} -v_{i+3} v_{i+4} \;,\qquad i = 1,\ldots,5 \label{a_def}
\end{align}
where we assume cyclicity of $v$-variable labels, see \cref{eq:kinematics},
and for the sake of presentation we split 31 letters $\{ W_i \}_{i=1}^{31}$ of the alphabet in the orbits of the cyclic group $\mathbb{Z}_5$,
\begin{equation} \label{pent_alph}
  \begin{aligned}
    & W_1 = v_1, \; W_2= v_2, \; W_3= v_3, \; W_4= v_4, \; W_5= v_5, \\[1em]
    &  W_6= v_3+v_4, \; W_7=v_4+v_5, \; W_8= v_1+v_5, \; W_9= v_1+v_2, \; W_{10}= v_2+v_3, \\
    & W_{11}= v_1-v_4, \; W_{12}=
    v_2-v_5, \; W_{13}= v_3-v_1, \; W_{14}= v_4-v_2, \; W_{15}= v_5-v_3, \\[1em]
    & W_{16}=
    v_1+v_2-v_4, \; W_{17}= v_2+v_3-v_5, \; W_{18}= v_3+v_4-v_1, \\ 
    & W_{19}= v_4+v_5-v_2, \; W_{20}= v_1+v_5-v_3, \\[1em]
    &  W_{21}= v_3+v_4  -v_1-v_2 , \; W_{22}=
    v_4+v_5 -v_2-v_3 , \; W_{23}= v_1+v_5-v_3-v_4, \\ 
    &W_{24}= v_1+v_2-v_4-v_5, \; W_{25}=
    v_2+v_3-v_1-v_5, \\[1em]
    &  W_{26}= \frac{a_{1}-\ep_5}{a_{1}+\ep_5},\; W_{27}= \frac{a_{2}-\ep_5}{a_{2}+\ep_5}, \; 
      W_{28}= \frac{a_{3}-\ep_5}{a_{3}+\ep_5}, \; W_{29}= \frac{a_{4}-\ep_5}{a_{4}+\ep_5},\;  W_{30}= \frac{a_{5}-\ep_5}{a_{5}+\ep_5}, \\[1em]
    & W_{31}= \ep_5  \,. \\[1em]
  \end{aligned}
\end{equation}
In the physical scattering region $\Delta < 0$ and $\ep_5$ is pure imaginary, see \cref{sect_phys_reg}. 
Since the Mandelstam variables are real, the letters $\{W_{25+i}\}_{i=1}^{5}$ are pure phases,
\begin{align}
W_{25+i}(X) = \exp( \ii\varphi_i(X) ) \quad \text{and} \quad \varphi_i(X) \;\; \text{is real}\,, \qquad i = 1 ,\ldots,5 \,.  \label{phase}  
\end{align}
We call letters parity-even or parity-odd according to parity-conjugation properties of their $\dd{\log}$-forms,
\begin{subequations}
  \begin{align}
  & \dd\log W_i \xrightarrow{\ep_5 \to -\ep_5} \dd\log W_i   \, , \quad & i &= 1 ,\ldots,25,31 \,, \\
  & \dd\log W_i \xrightarrow{\ep_5 \to -\ep_5} -\dd\log W_i  \, , \quad & i &= 26 ,\ldots,30 \,.
  \end{align}
\end{subequations}
Thus, $\{ W_i \}_{i = 1}^{25}$ and $W_{31}$ are {\it parity-even} and $\{W_{i}\}_{i=26}^{30}$ are {\it parity-odd}.
Note that, since $\ep_5$ is imaginary in the physical region, the parity-conjugation is equivalent to the complex conjugation in this region.

Working in the non-planar sector we have to consider all ${\cal S}_5$ permutations of the external momenta $p^\mu_1, \ldots,p^\mu_5$. 
The alphabet is closed under this action, which induces representation of ${\cal S}_5$ in the space of the letters. 
The decomposition in the irreducible representations of ${\cal S}_5$ can be found in \cite{Chicherin:2017dob}. 
Here we prefer to work with reducible representations, and we summarize the action of ${\cal S}_5$ on the letters:
\begin{itemize}
\item
The sets of ten even letters $\{ W_i \}_{i = 1}^5 \cup \{ W_i \}_{i =16}^{20}$ and fifteen even letters  $\{ W_i \}_{i = 6}^{15} \cup \{ W_i \}_{i =21}^{25}$ are closed under ${\cal S}_5$. 
The permutations map letters into each other up to sign, i.e.\ $W_i \to W_j$ or $W_i \to - W_j$.
\item
The five parity-odd letters $\{ W_{i} \}_{i= 26}^{30}$ transform non-linearly under ${\cal S}_5$.  
The transformations look like: $W_i \to W_j$, $W_i \to W_{j}^{-1}$, $W_{i} \to W_{j}W_{j+1}$, $W_{i} \to W_{j}^{-1}W_{j+1}^{-1}$,
$W_{i} \to W_{j}W_{j+1}W_{j+3}^{-1}$ and $W_{i} \to W_{j}^{-1}W_{j+1}^{-1}W_{j+3}$ where all indices run cyclically over $26,\ldots,30$,
so only the five parity-odd letters appear in the transformation rules.  
\item
$W_{31}$ is mapped to itself up to sign, i.e.\ $W_{31} \to W_{31}$ or $W_{31} \to -W_{31}$ depending on the signature of ${\cal S}_5$ permutation. 
\item Considering $\dd{\log}$-forms of the letters, the previous transformations simplify. 
For the parity-even letters, the action of ${\cal S}_5$ on $\{ \dd\log W_i \}_{i=1}^5 \cup \{ \dd\log W_i \}_{i=16}^{20}$,  $\{ \dd\log W_i \}_{i=6}^{15} \cup \{ \dd\log W_i \}_{i=21}^{25}$ and $\{ \dd\log W_{31} \}$ is by permutations, and for the parity-odd letters, $\{ \dd\log W_i \}_{i = 26}^{30}$ transform linearly.

\end{itemize}

\section{Physical region geometry}
\label{App_conv}

The analyticity region $P_0$ \p{s12channel} is not convex, but any line segment from $X_0$ \p{B} to any $X \in P_0$ lies inside $P_0$. 
Let us prove this statement. 
We are going to show that any ray from $X_0$ crosses the boundary of $P_0$ only once, i.e. it cannot enter back inside $P_0$. 
We need to consider only $\Delta = 0$ boundary of $P_0$. 
For other boundaries of $P_0$, i.e. $s_{ij} = 0$ with $i,j=1,2$ or $i,j = 3,4,5$ or $i=1,2$ and $j=3,4,5$, the statement is obvious. 
Let us choose an arbitrary $X_1$ on the boundary of $P_0$,
\begin{align}
\Delta(X_1) = 0 \;, \quad s_{12},s_{34},s_{35},s_{45}|_{X_1} \geq 0
\;, \quad  s_{13},s_{14},s_{15},s_{23},s_{24},s_{25}|_{X_1} \leq 0 \,, \label{X1bound}
\end{align}
and connect it with $X_0$ by the line segment $\gamma = [X_0 ; X_1]$ parametrized by $0 \leq t \leq 1$. 
We need to prove that $\Delta(t) \equiv \Delta(\gamma(t)) <0$ at $1-\varepsilon<t<1$ for small positive $\varepsilon$.
Since $\Delta$ is a degree-4 polynomial in the Mandelstam variables, it is enough to show that 
none of the following four inequalities is compatible with \p{X1bound},  
\begin{subequations}
  \begin{align}
  & \Delta'(1) < 0 \;\; \text{at} \;\; \Delta(1) = 0 \;; \\
  &   \Delta''(1) > 0 \;\; \text{at} \;\;  \Delta'(1) = 0 \;,\; \Delta(1) = 0 \;;\\
  &   \Delta'''(1) < 0 \;\; \text{at} \;\; \Delta''(1) = 0 \;,\;  \Delta'(1) = 0 \;,\; \Delta(1) = 0 \;;\\
  &   \Delta''''(1) > 0 \;\; \text{at} \;\; \Delta'''(1) = 0 \;,\; \Delta''(1) = 0 \;,\;  \Delta'(1) = 0 \;,\; \Delta(1) = 0 \,.
  \end{align}
\end{subequations}
This can be verified using a computer algebra system. 
Extra simplification of the inequalities is achieved by fixing value of one of $s_{ij}$ (since all expressions are homogeneous polynomials in the Mandelstam invariants)
and by using $\Delta^{(k-1)}(1) = 0$ to lower the degree of $\Delta^{(k)}(1)$.

\bibliography{main}

\end{document}